\documentclass[12pt]{aastex61}

\shorttitle{RADIO BURSTS IN SOLAR ERUPTIVE FLARES} \shortauthors{Karlick\'y}
\begin{document}

\title{RADIO BURSTS OBSERVED DURING SOLAR ERUPTIVE FLARES AND THEIR SCHEMATIC SUMMARY}

\author[0000-0002-3963-8701]{Marian Karlick\'y}
\affil{Astronomical Institute of the Academy of Sciences of the
Czech Republic, CZ-25165 Ond\v{r}ejov, Czech Republic}
 \email{karlicky@asu.cas.cz}

\begin{abstract}
In this review we summarize results of our analysis of the observations of
solar eruptive flares made by the Ond\v{r}ejov radiospectrograph for more than
twenty years. We also present some Potsdam-Tremsdorf radio spectra from our
common studies. Considering a 3-dimensional model of eruptive flares together
with the results of our magnetohydrodynamic and particle-in-cell simulations we
show an importance of decimetric radio bursts for understanding of plasma
processes in eruptive flares. We present drifting pulsation structures as
signatures of plasmoids, an unusual zebra pattern in the very early flare
stage, narrowband dm-spikes as the bursts generated in the reconnection plasma
outflows, radio bursts indicating a merging of plasmoids, pair of decimetric
type III bursts indicating the electron beams propagating upwards and downwards
in the solar atmosphere from the acceleration site, and a special decimetric
type III burst formed probably around the plasmoid. We present unusual radio
bursts connected with the rising magnetic rope at the very beginning of
eruptive flares. Furthermore, based on the analysis of decimetric continua we
estimated the level of the plasma turbulence in a vicinity of the flare
termination shock. Interpretations of all these bursts are based on models and
time coincidences with observations in X-ray, UV and optical ranges; in most
cases an information about positions of these radio sources is missing. To show
an importance of positional information, we present a rare example of
observations, where the drifting pulsation structure was observed
simultaneously with the observations made by the EOVSA radiointerferometer. All
the presented bursts are then summarized in a new scheme of bursts and compared
with the schema commonly used.
\end{abstract}

\keywords{Sun: flares -- Sun: radio radiation}

\section{INTRODUCTION}

Solar flares are the most powerful events in the the solar system, releasing
the energy up to 10$^{32}$ erg in tens of minutes. The strongest flares are
associated with coronal mass ejections and acceleration of particles into
interplanetary space. From physical point of view solar flares are explosive
phenomena in the solar atmosphere, in which the energy accumulated in the
magnetic field and electric currents is rapidly transformed into plasma
heating, plasma flows, accelerated particles and emission in a broad range of
electromagnetic waves: from radio, through optical, UV, X-rays to gamma-rays.
For more details, see reviews by in
\cite{2002A&ARv..10..313P,2002SSRv..101....1A,2008A&ARv..16..155K,2009AdSpR..43..739S,2011SSRv..159...19F,2016SSRv..200...75N}.

Generally, solar flares can be divided into eruptive flares and confined flares
\citep{1992LNP...399.....S}. Note that the eruptive flares are sometimes called
the LDE (long duration event) flares or two-ribbon flares. The eruptive flares
belong to the most powerful classified in GOES classification as M and X
flares. These flares are described by the "standard" CSHKP flare model
\citep{1964NASSP..50..451C,1966Natur.211..695S,1974SoPh...34..323H,1976SoPh...50...85K},
or by its generalized 3-dimensional model proposed by
\citep{2012A&A...543A.110A,2014ApJ...788...60J}.

Because the eruptive flares influence the whole heliosphere, including the
space around the Earth, these flares are of great interest for the space
weather programs. They are monitored by many instruments on board of satellites
and ground based observing stations in a broad range of emissions. These
emissions are produced by various emission mechanisms, e.g., the bremsstrahlung
for X-rays, transitions of electrons between energetic atomic levels for UV and
optical emissions, gyro-synchrotron and plasma emission mechanisms for radio
emission and so on. Among these mechanisms the plasma emission mechanism is
special one because it is directly connected with the plasma processes in
flares. Thus, solar decimetric and metric radio bursts, which are generated by
this emission mechanism, give us a direct insight into flare processes.
Unfortunately the bursts, especially in decimetric range, which is of high
importance for diagnostics of eruptive flares, are usually only observed
spectroscopically. Observations with spatial information about their sources
are rare. Nevertheless, as will be shown in the following, owing to a high-time
resolution of dynamical radio spectra, important results about plasma processes
in solar flares can be obtained.

This review paper presents examples of the radio bursts observed in decimetric
range during the eruptive flares by the Ond\v{r}ejov radiospectrographs
\citep{1993SoPh..147..203J,2008SoPh..253...95J}. Examples were selected from
numerous radio spectra recorded for more than twenty years. The paper is
organized as follows: After describing a model of the eruptive flare we present
a typical radio spectrum of the eruptive flare with type III and II bursts,
drifting pulsation structure and high-frequency continua. Then we show unique
drifting bursts in connection with the rising magnetic rope, unusual zebra
pattern and slowly positively drifting bursts at the very beginning of flares.
We show radio bursts connected with plasmoids, narrowband dm-spikes, a pair of
normal and reverse drift type III bursts and unusual type III burst connected
probably with a plasmoid. We also present the power spectral analysis of
decimetric continuum and its new model. Then we show one example of the
drifting pulsation structure observed simultaneously with the EOVSA spatial
observations. The presented bursts are interpreted within a 3-dimensional model
of eruptive flares and results of numerical simulations of plasma processes in
solar flares. We summarize all presented bursts into the schema which completes
the "standard" schema of radio bursts in solar flares shown, e.g., in
\cite{1979itsr.book.....K}. The paper ends with conclusions and statements
expressing an importance of new instruments which would yield not only
spectral, but combined spectral and spatial information about observed radio
bursts.

\section{MODEL OF THE ERUPTIVE FLARE}

\begin{figure}
\begin{center}
\includegraphics*[width=16.0cm]{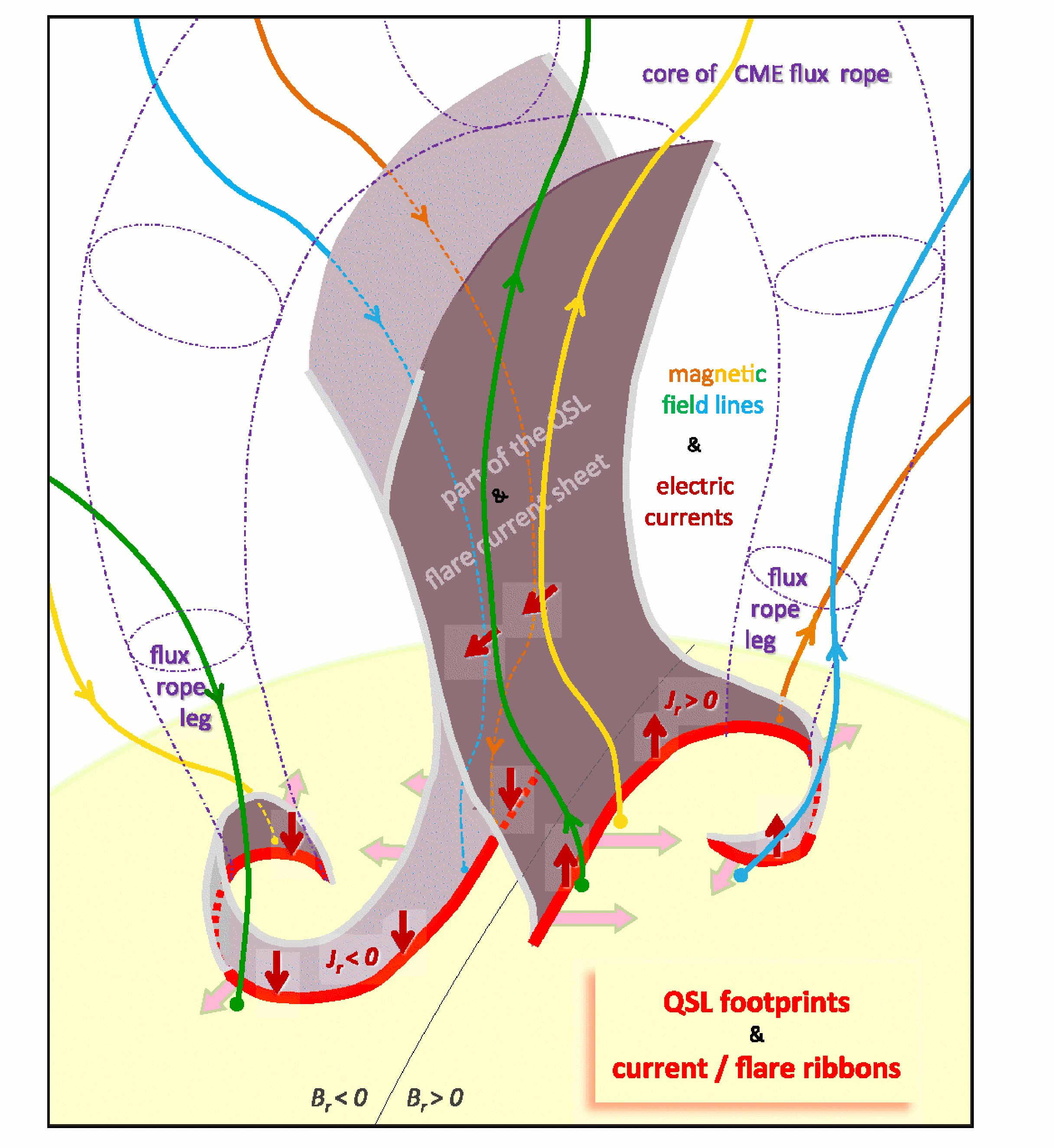}
\end{center}
    \caption{The 3D model of eruptive solar flares \citep{2014ApJ...788...60J}.}
\label{fig1}
\end{figure}

A sketch of the 3-dimensional model of the eruptive flare is shown in
Figure~\ref{fig1}. In this model the eruptive flare can be described as follows
\citep{2002A&ARv..10..313P,2002SSRv..101....1A,2008A&ARv..16..155K,2009AdSpR..43..739S,
2011SSRv..159...19F,2016SSRv..200...75N,2012A&A...543A.110A,2014ApJ...788...60J}:

First, along the neutral line between magnetic polarities in the active region,
a magnetic rope (i.e. current-carrying loop) is formed owing to shear and
vortex plasma flows at the photospheric level. At its bottom part a cold and
dense plasma condensates into the filament, which is dark in contrast to the
surrounding bright chromosphere. Then this magnetic rope together with the
filament becomes unstable due to torus or kink instabilities or external
perturbations (particle beams, waves or shocks)
\citep{2006PhRvL..96y5002K,2010SoPh..266...91K,1989SoPh..124..319K,1997A&A...326.1252O}.
Another possibility is that the upper part of the magnetic rope interacts with
the above-lying magnetic field lines and through magnetic reconnection a
stabilizing magnetic force decreases \citep{2011ApJ...730...57A}. As a result
of all the mentioned processes the magnetic rope moves upwards, but its ends
are still anchored in dense layers of the solar atmosphere, forming thus a
growing current-carrying loop. Although an enormous electric current can flow
through this magnetic rope (up to 10$^{12}$ A), this current does not dissipate
here, because the electric current density is low. The situation is different
below the rising rope, where the current sheet is formed step by step
\citep{2018ApJ...853L..18Y}. It is assumed that when this current sheet becomes
sufficiently narrow and the current density inside this sheet sufficiently
high, then the plasmoid instability starts and magnetic reconnection follows
\citep{2017ApJ...849...75H}. A plasma together with magnetic field lines is
sucked from both sides of the current sheet into the X-point region and
accelerated to the Alfv\'en speed in vertical (upward and downward) plasma
outflows. The upward oriented outflow even accelerates the rising magnetic
rope, while the downward oriented outflow is stopped above the flare loop
arcade, where the so called termination shock can be formed
\citep{2004ApJ...615..526A}. At the X-point within the magnetic reconnection
region a strong electric field is generated, which accelerates electrons and
ions to high energies \citep{2008ApJ...674.1211K,2014PhPl...21i2304D}. These
particles propagate along the magnetic field lines. Some of them propagate
upwards in the solar atmosphere, generating radio emission by the plasma
emission mechanism, e.g. type III radio bursts \citep{1985srph.book..289S}. The
electrons propagating downwards generate radio continua (at frequencies above
about $\sim$ 2 GHz)  by the gyro-synchrotron mechanism
\citep{1979itsr.book.....K} and the hard X-ray emission by bremsstrahlung
\citep{1971SoPh...18..489B}. Simultaneously, electrons in the whole flare
generate various fine structures in the radio spectrum via the plasma emission
mechanisms \citep{2011fssr.book.....C}. The superthermal ions, especially
protons, propagating downwards are sources of gamma-rays
\citep{1996SoPh..166..107A}. During the magnetic reconnection process the flare
plasma is strongly heated, which leads to an increase of the soft X-ray
emission. The whole magnetic flare structure expands in the process called
coronal mass ejection (CME) and magnetohydrodynamic shocks are produced. Such
shocks then generate type II radio bursts \citep{1985srph.book..333N}.
Furthermore, associated waves propagate along the solar surface as Moreton and
EIT (UV) waves \citep{1960AJ.....65U.494M,1998GeoRL..25.2465T}. During some
flares even large amplitude loop oscillations and seismic waves have been
detected \citep{2016SSRv..200...75N,1998Natur.393..317K}.

\section{EXAMPLES OF RADIO BURSTS AND THEIR INTERPRETATIONS}

In the following we present examples of radio bursts observed during eruptive
flares. They are interpreted based on the 3-dimensional model of eruptive
flares and corresponding numerical simulations.

\subsection{Radio bursts observed during the 12 April 2001 flare}

\begin{figure}
\begin{center}
\includegraphics*[width=8.0cm]{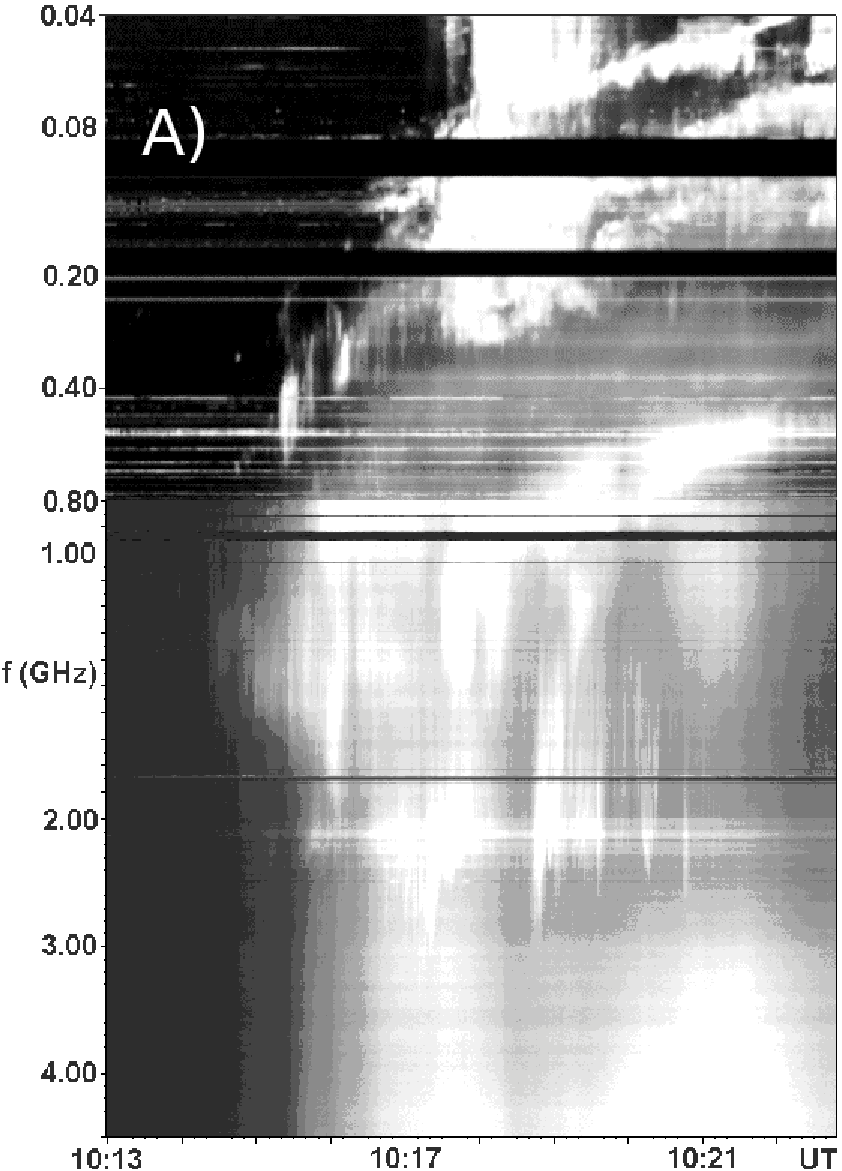}
\includegraphics*[width=8.0cm]{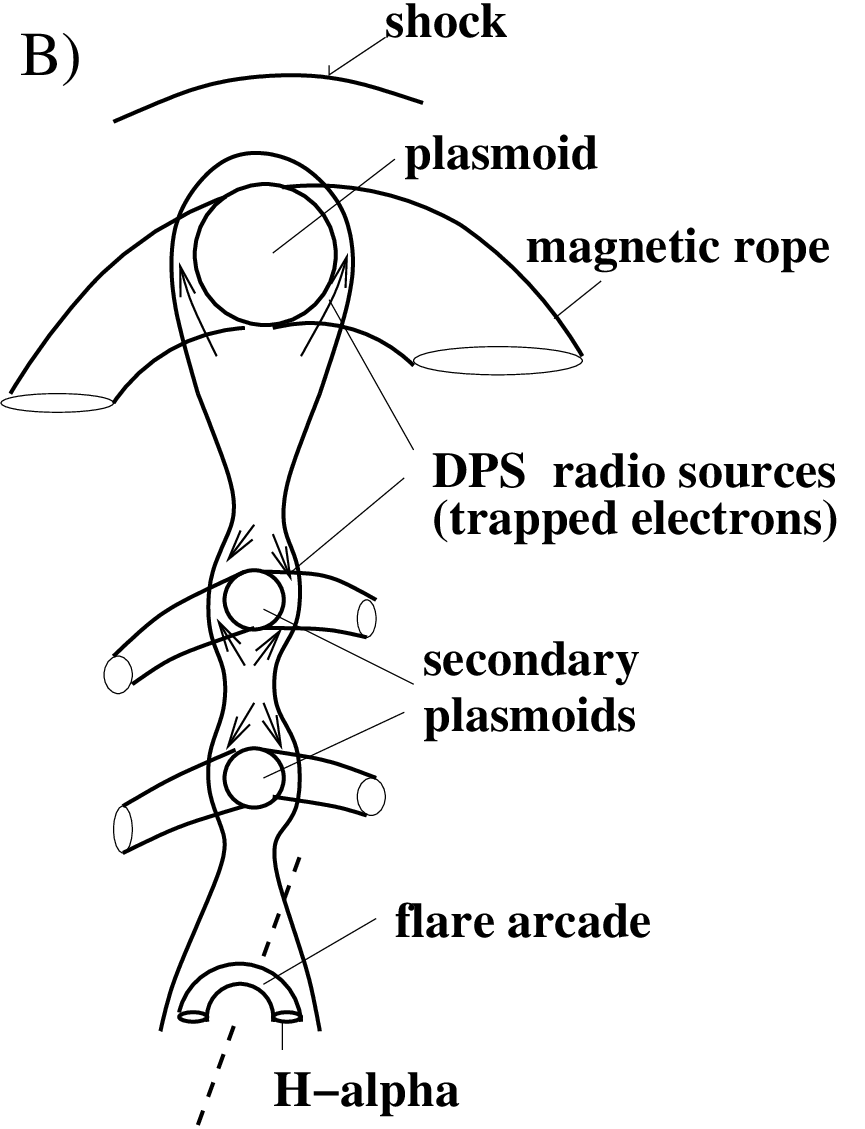}
\end{center}
    \caption{A) The 0.04--4.5 GHz radio spectrum observed during the 12 April 2001
flare by the Potsdam-Tremsdorf (0.04--0.8 GHz range) and the
Ond\v{r}ejov radiospectrographs (0.8--4.5 GHz), showing the drifting pulsating structure (DPS) at 10:17:20 --
10:22:00 UT in the 0.45--1.5 GHz range and the type II radio burst (generated
by a shock wave) at 10:17 -- 10:33 UT in the 0.04--0.3 GHz range. B) Flare
scenario: The rising magnetic rope (the uppermost plasmoid) generates underneath a vertical
current sheet, where further plasmoids (producing DPSs) are generated due
to the tearing-mode instability. Above this main rope, the shock, producing type
II radio burst, is generated; compare this scenario with the radio spectrum in
the left part of the figure.}
\label{fig2}
\end{figure}

Figure~\ref{fig2}A shows a typical radio spectrum of the eruptive flare. This
spectrum was recorded during the 12 April 2001 flare, classified as X2.0 flare
which started at 09:39 UT and ended at 10:49 UT.  At frequencies above about 3
GHz this spectrum shows the continuum emission (type IV radio burst), which is
believed to be generated by the gyro-synchrotron emission mechanism. On the
other hand, below 3 GHz the observed bursts are produced by the plasma emission
mechanism due to their fine structures. In this case, the square of their
frequency is proportional the the electron plasma density in their sources.
Because the solar atmosphere is gravitationally stratified, i.e., the plasma
density decreases increasing the height in the solar atmosphere, a frequency
distribution of bursts in this spectrum roughly corresponds to the distribution
of their sources in heights. Thus, we can see that type III bursts, observed as
rapidly drifting in the 0.04 -- 0.3 GHz range at about 10:17 -- 10:20 UT and
generated by electron beams (propagating into the upper corona), are at higher
altitude than that of the drifting pulsating structure (DPS) observed  at
10:14:30 -- 10:22:00 UT in the 0.45 -- 2.0 GHz range. Furthermore, the spectrum
shows two branches of type II radio burst observed at 10:15 -- 10:23 UT in the
0.04 -- 0.6 GHz range and generated by the flare shock. To show correspondence
of these bursts with the 3-D model of the eruptive flare, in Figure~\ref{fig2}B
a model scenario is shown. The rising magnetic rope (the uppermost plasmoid)
generates underneath a vertical current sheet, where secondary plasmoids and
flare loop are generated due to the plasmoid instability. Above the magnetic
rope the shock is produced owing to its super-alfvenic velocity. While the high
frequency continuum is generated at the bottom part of this structure, at flare
loops, shock generates type II burst on fundamental and harmonic plasma
frequency. Electrons are accelerated at X-magnetic field points located between
plasmoids. Some electrons escape from this structure and generate type III
bursts, Other accelerated electrons are trapped in the plasmoid of the magnetic
rope or in secondary plasmoids. Here, the electrons generate the plasma waves
that are then transformed to the electromagnetic (radio) waves observed as
DPSs. The density in the plasmoid is in some range, which thus corresponds to
the bandwidth of DPS. The plasmoids move upwards or downwards along the flare
current sheet with velocities that are smaller than local Alfv\'en speed. In
the radio spectrum the plasmoid motion is expressed by the frequency drift of
DPS. Pulses in DPS are owing to a quasi-periodic regime of the electron
acceleration. For more details about the drifting pulsation structure and its
model, see \cite{2000A&A...360..715K,2008SoPh..253..173B}. As concerns to more
details about this radio event, see \cite{2006A&A...460..865M}.

\subsection{Radio burst observed at the beginning of the 24 September 2001 flare}
\begin{figure}
\begin{center}
\includegraphics*[width=18.0cm]{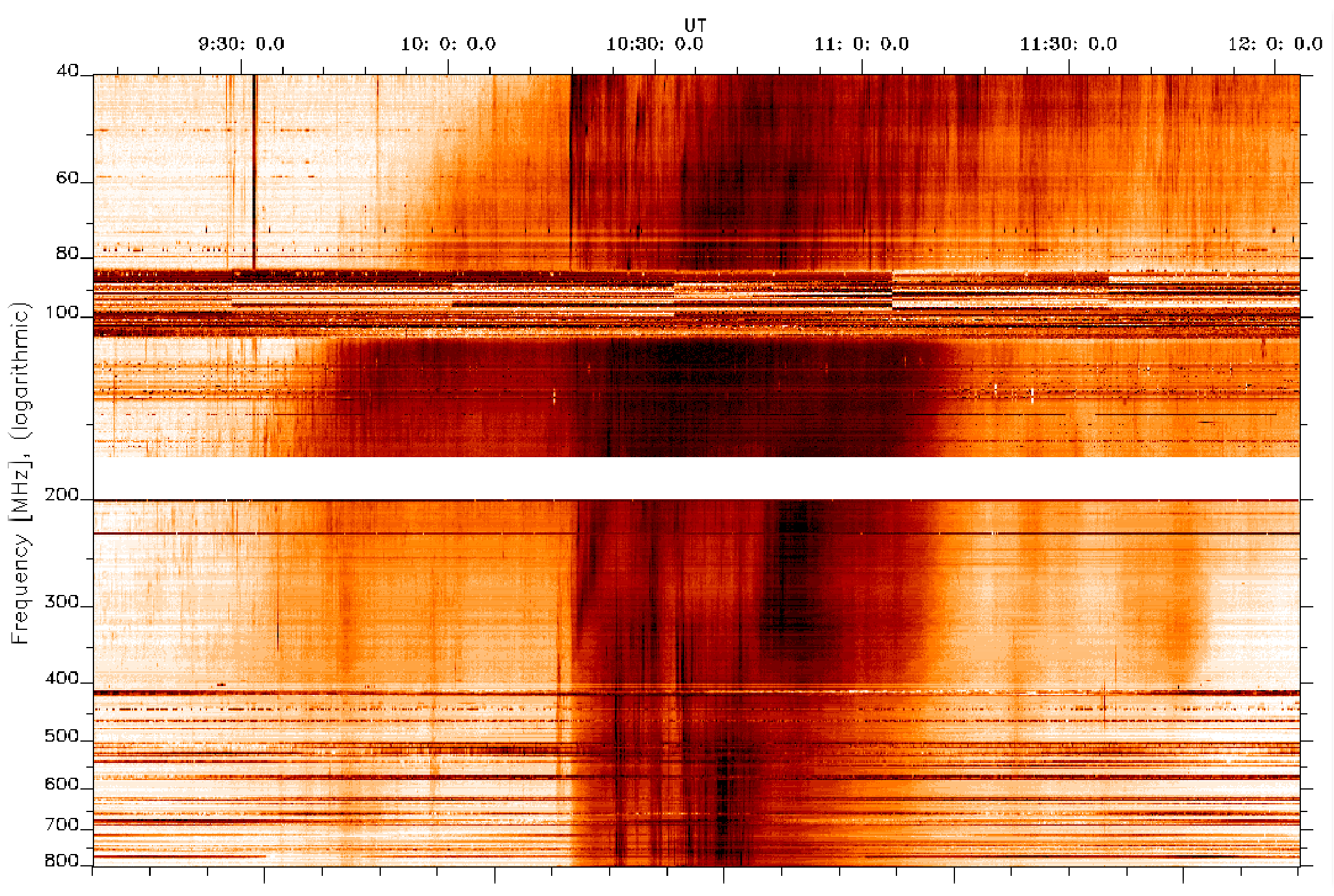}
\end{center}
    \caption{The 40-800 MHz Potsdam-Tremsdorf radio spectrum observed at 09:00-12:00 UT on September 24, 2001 (courtesy Dr. A. Klassen).}
\label{fig3}
\end{figure}

At the very beginning of the 24 September 2001 flare (X2.6, start 09:20 UT and
end 11:39 UT), just before a series of type III bursts (starting at 10:18 UT
and indicating the flare impulsive phase), a very unusual burst was observed at
09:30 -- 10:18 in the 800 -- 40 MHz range (Figure~\ref{fig3}). It happened in
the phase when the magnetic loop rises \citep{2003A&A...399.1159F}. Therefore,
we call this burst as the rope rising continuum. It looks as some continuum
superimposed by some weak type I-like bursts. Its starting boundary drifts from
higher to lower frequencies. Assuming the plasma emission mechanism of this
burst and the emission on the fundamental frequency then the frequency drift of
this burst in the model of the solar atmosphere \citep{2002SSRv..101....1A}
corresponds to the upwards velocity 280 km s$^{-1}$.  This velocity is in the
range of the velocities of rising magnetic rope, therefore we think that this
burst in generated on the outer boundary of the rising rope, where the rope
interacts with the above-lying magnetic field or inside the rope due to some
inner-rope magnetic reconnection. Moreover, we think that type III bursts
started just when below the rising magnetic rope a sufficiently narrow current
sheet is formed and the flare magnetic reconnection starts. Such a burst is
very rare. We found a similar burst only at the very beginning of the 7 June
2011 flare \citep{2020ApJ...888...18K}.

\subsection{Radio bursts observed at the beginning of the 6 September 2017 flare}

\begin{figure}
\begin{center}
\includegraphics*[width=18.0cm]{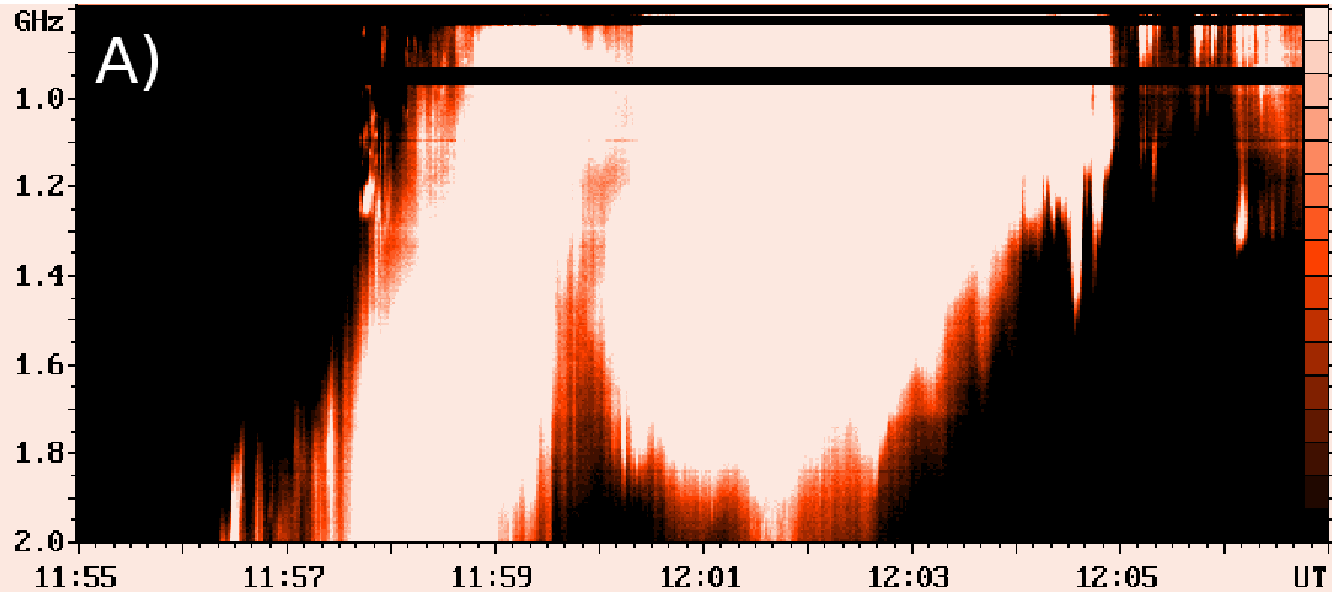}
\includegraphics*[width=18.0cm]{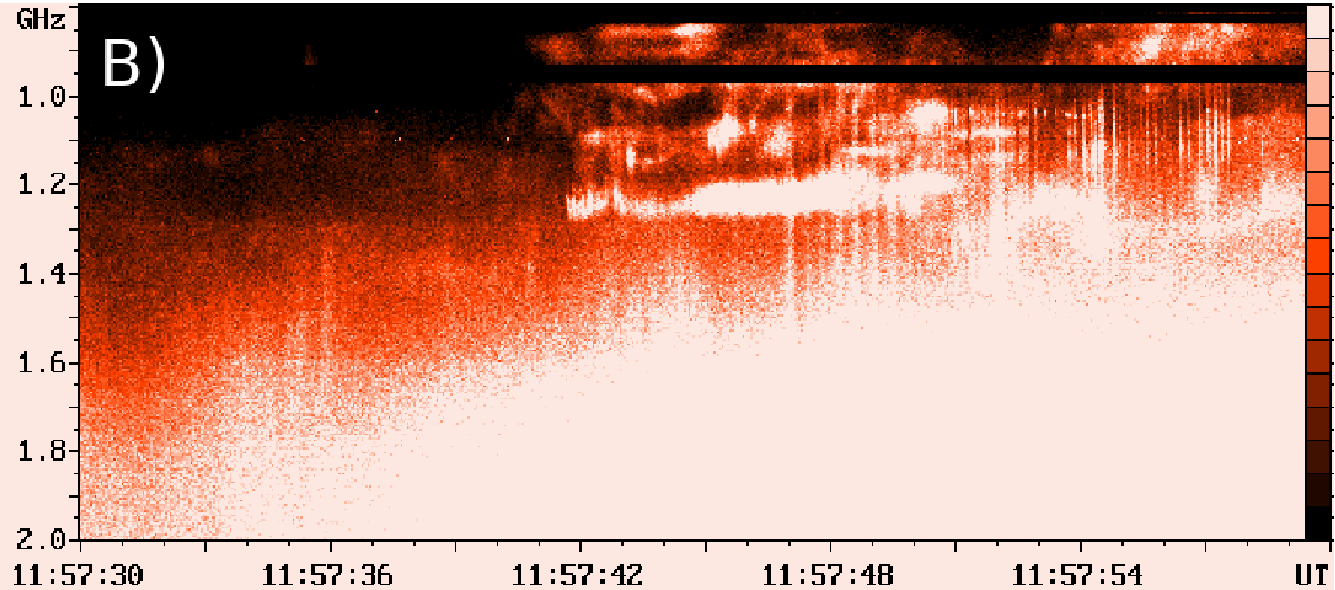}
\end{center}
    \caption{A) Radio spectrum of the 6 September 2017 flare observed by the
Ond\v{r}ejov radiospectrograph. B) The detail showing an unusual zebra-like burst.}
\label{fig4}
\end{figure}

The 6 September 2017 flare was classified as X9.3 with its start at 11:53 UT
and end at 12:10 UT. The flare was so strong that the impact of its radio
emission has been found at the frequency 1.2 GHz of Global Navigation Satellite
System \citep{2018AGUFMSA34A..04S}. For more details about this flare, see
\cite{2018ApJ...869...69M} and references therein.   The radio spectrum at the
beginning of this flare is shown in Figure~\ref{fig4}A. As seen here, the whole
radio emission is rapidly drifting to lower frequencies. The most remarkable
aspect of this radio burst is the zebra-like structure observed at the starting
boundary of this globally drifting burst (Figure~\ref{fig4}B). Owing to the
flare phase of its observation and relatively broadband zebra stripes it
differs from normal zebras that are usually observed in later flare phases.
Although we cannot exclude that this zebra-like structure is generated by some
of mechanisms considered for normal zebras \citep{2011fssr.book.....C}, we
propose that at this flare phase the zebra-like structure can be the radio
emission from four plasmoids distributed in the vertical flare current sheet.
If so then the emission mechanism generating this zebra-like structure would be
the same as in DPSs \citep{2000A&A...360..715K}.

\subsection{Slowly positively drifting bursts observed in the 2 April 2022 flare}

\begin{figure}
\begin{center}
\includegraphics*[width=14.0cm]{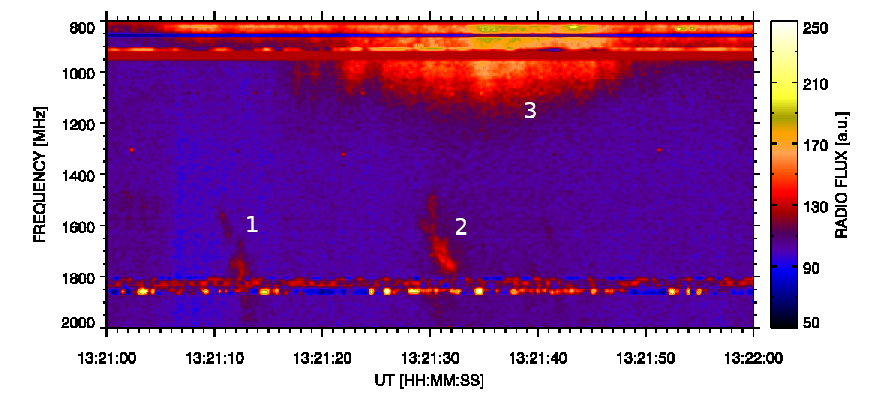}
\includegraphics*[width=14.0cm]{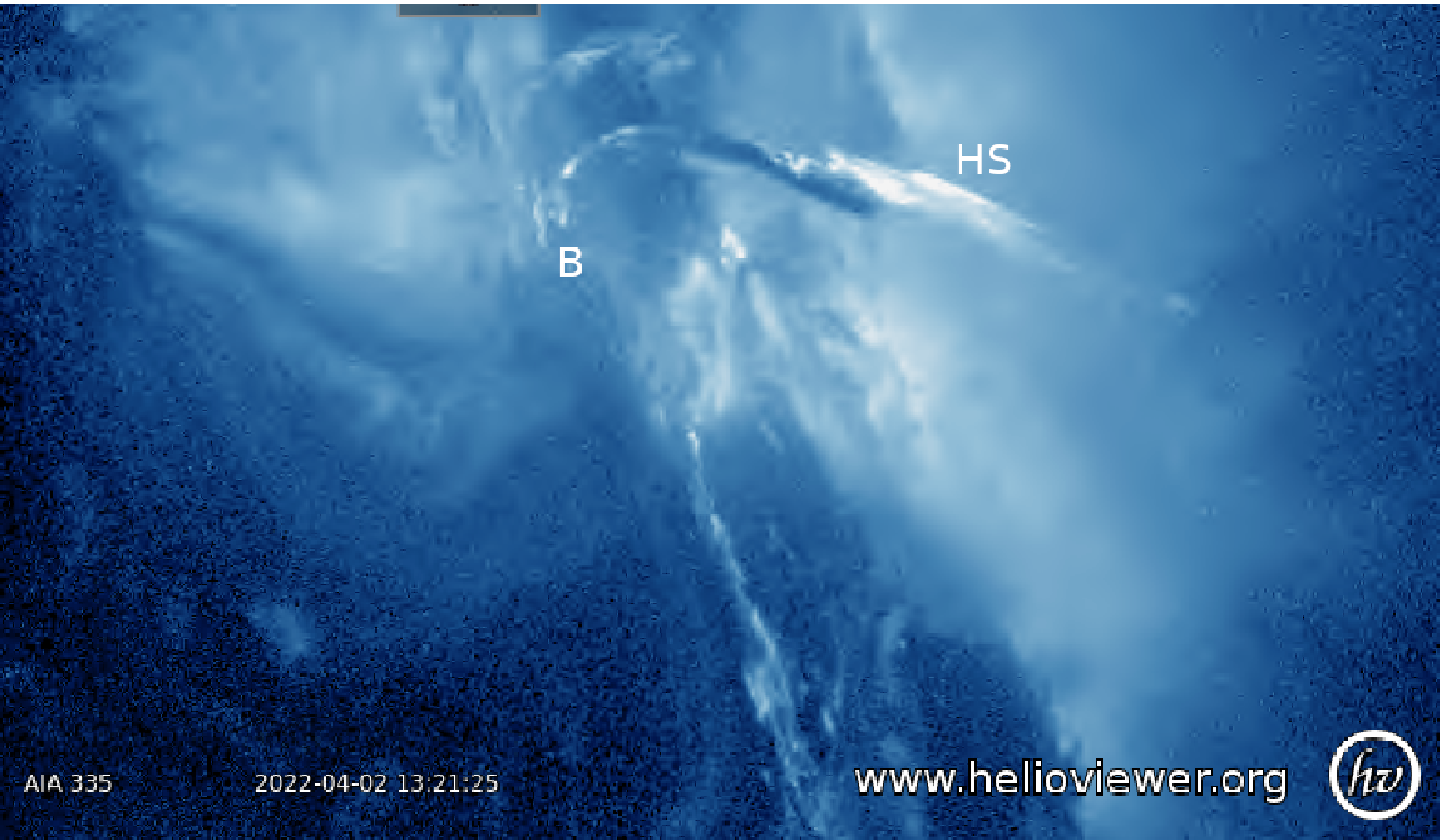}
\end{center}
    \caption{Upper: Radio spectrum at the very beginning of the 2 April 2022 flare with
    the slowly positively drifting bursts (1 and 2), 3 means continuum. Bottom:
    The associated AIA/SDO 335 \AA~image, observed at 13:21:25 UT, showing a bright helical structure (HS) inside
    the rising magnetic rope. B means the brightening at the footpoint of the rising magnetic rope.}
\label{fig5}
\end{figure}

Figure~\ref{fig5} upper part shows the radio spectrum observed at the very
beginning of the eruptive flare, where two slowly positively drifting bursts (1
and 2) can be seen. Their frequency drift is about 100 MHz s$^{-1}$. At the
same time the AIA/SDO image (Figure~\ref{fig5} bottom part) shows a bright
helical structure (HS) as a part of the rising magnetic rope. In accordance
with \cite{2020ApJ...905..111Z} we interpret these slowly positively drifting
bursts as generated by the particle beams propagating in the bright helical
structure. Owing to their propagation along the helical trajectory, that is
even deflected from the vertical direction in the gravitationally stratified
atmosphere, the frequency drift of these bursts is much smaller compared with
the typical frequency drift of the reverse type III bursts in this frequency
range ($\sim$ 1 GHz s$^{-1}$). This interpretation is supported by a
brightening (B) at the location, where the magnetic rope is anchored.  The
bright helical structure as well as slowly positively drifting bursts show a
flaring activity in the rising magnetic rope. Note, that such an activity is
not considered in the standard CSHKP flare model.

\subsection{Drifting pulsation structure observed during the 18 August 1998 flare}

\begin{figure}
\begin{center}
\includegraphics*[width=16.0cm]{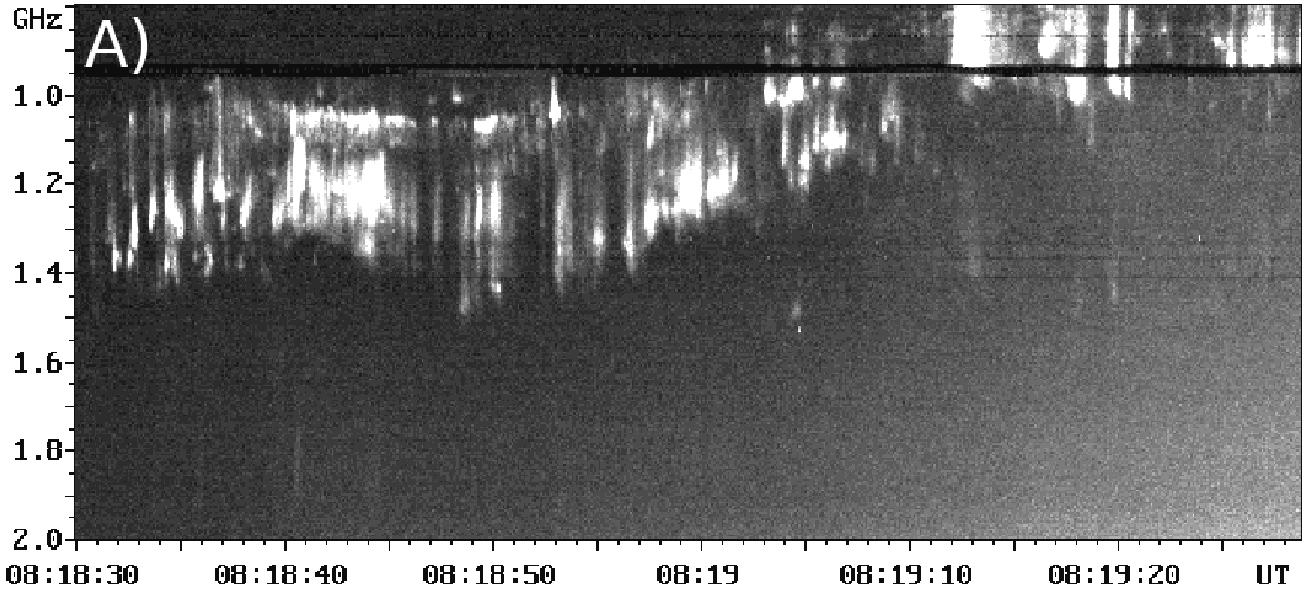}
\includegraphics*[width=18.0cm]{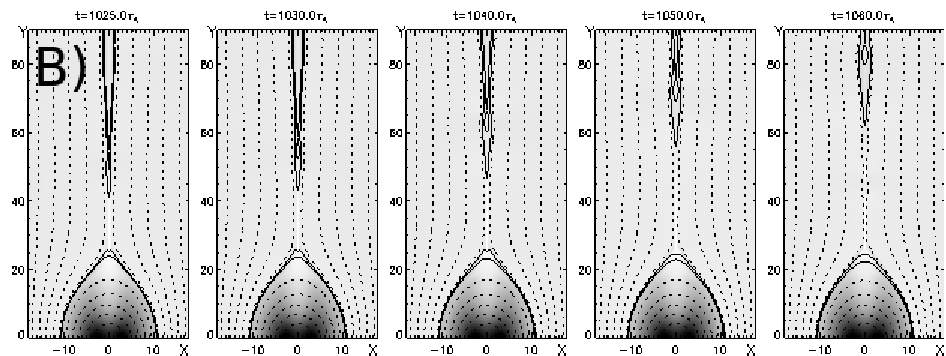}
\end{center}
    \caption{A) Radio spectrum of the 18 August 1998 flare observed by the
Ond\v{r}ejov radiospectrograph showing the drifting pulsation structure (DPS). B)
Results of numerical simulations showing formation of the plasmoid that moves upwards
in the solar atmosphere, see \cite{2008A&A...477..649B}.}
\label{fig6}
\end{figure}

Drifting pulsation structures are commonly observed at the very beginning of
the eruptive flares in the decimetric range. One example is shown in
Figure~\ref{fig6}A that was observed during the X2.8 flare, which started at
08:18 UT and ended at 08:45 UT \citep{2002A&A...395..677K}. As seen in this
spectrum the frequency drift of DPS varies in time. But globally, the drift is
from higher to lower frequencies as in the most DPS cases. In agreement with
the interpretation of DPS in the paper by \cite{2000A&A...360..715K} we
interpret the drift variation by a change of the propagation direction of the
emitting plasmoid along the vertical flare current sheet and the global drift
as an upwards motion of the plasmoid. To support this interpretation we made
numerical simulations showing the upwards motion of the plasmoid in the current
sheet, see different evolution phases of the plasmoid in Figure~\ref{fig6}B.
For more details about these numerical simulations, see the paper by
\cite{2008A&A...477..649B}.

\subsection{Drifting pulsation structures and narrowband dm-spikes observed during the 28 March 2001 flare}

\begin{figure}
\begin{center}
\includegraphics*[width=8.0cm]{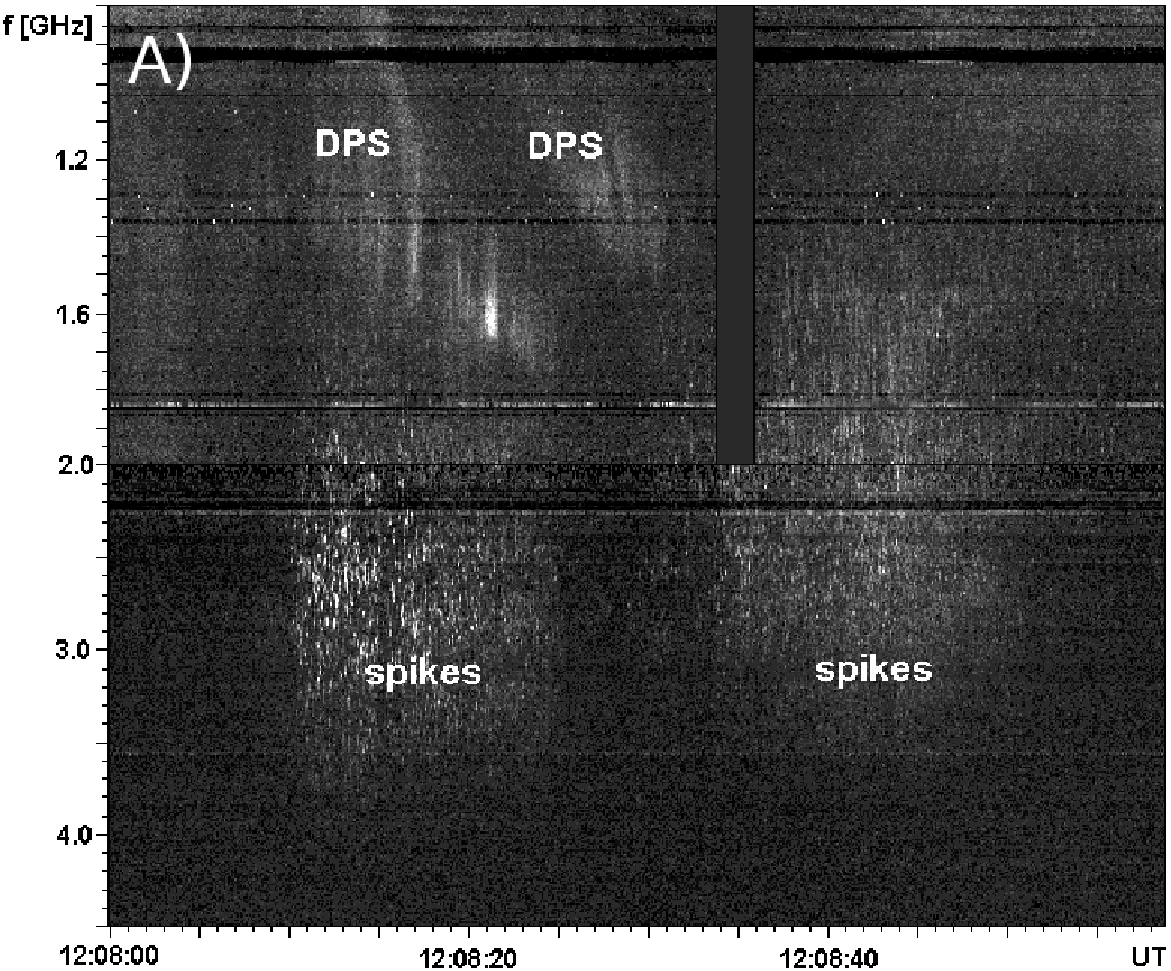}
\includegraphics*[width=8.0cm]{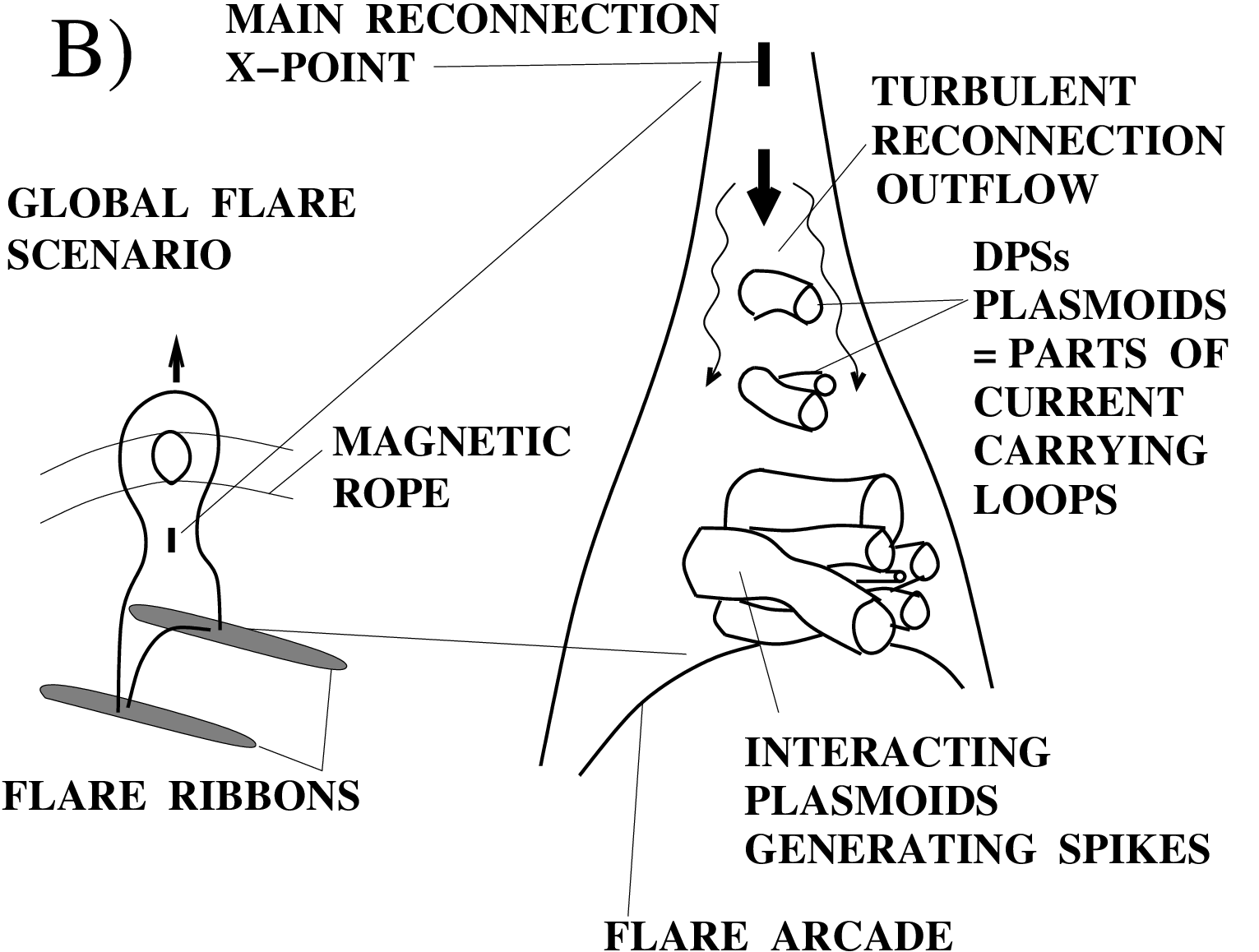}
\includegraphics*[width=18.0cm]{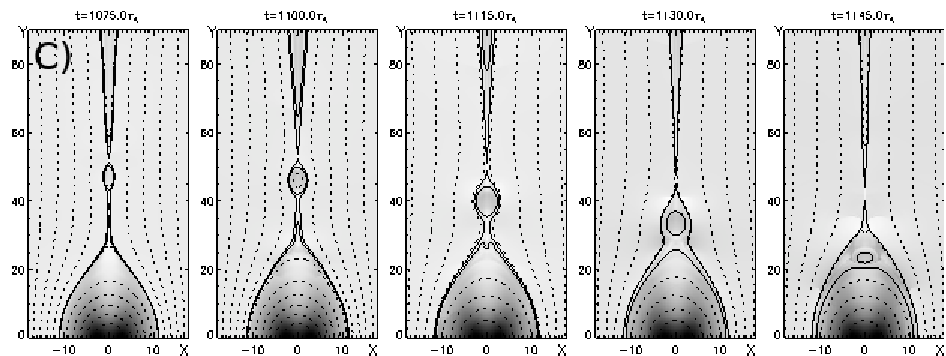}
\end{center}
    \caption{A) Radio spectrum of the 28 March 2001 flare observed by the
    Ond\v{r}ejov radiospectrographs showing two DPSs which drift towards the narrowband dm-spikes. B) Scenario of DPSs and dm-spikes generation.
    C) Results of numerical simulations showing formation of the plasmoid that moves downwards
in the solar atmosphere and interacts with the flare arcade, see \cite{2008A&A...477..649B}.}
\label{fig7}
\end{figure}

The numerical simulations presented in \cite{2008A&A...477..649B} also show
that the plasmoid can move downwards along the flare current sheet and
interacts with the underlying flare arcade (Figure~\ref{fig7}C). Therefore, we
searched for appropriate bursts and we found the radio spectrum of the 28 March
2001 flare (M4.3, start 12:08 UT and end 12:15 UT \citep{2003A&A...407.1115M}),
see Figure~\ref{fig7}A. Here, there are two DPSs with the positive frequency
drift that are followed by a cloud of the narrowband dm-spikes on higher
frequencies. To explain processes in these bursts a scenario is added in
Figure~\ref{fig7}B. Here, the plasmoids emitting DPSs move downwards as in the
numerical simulation and then they interact with the flare arcade. In the
papers by \cite{1996SoPh..168..375K,2000SoPh..195..165K} we found that the
Fourier spectra of the narrowband spikes are the power-law and their power-law
indices are close to the power-law index of the Kolmogorov spectra (-5/3).
Based on this finding we propose that during the interaction of plasmoids with
the flare arcade the plasmoids are fragmented. Thus, they form a cascade of
fragmented plasmoids, where each plasmoid emits as DPS. Because a cascade of
the fragmented plasmoids consists the plasmoids of different sizes then as a
result we observe a cloud of dm-spikes having different frequency bandwidths.

\subsection{Radio bursts observed during the 28 September 2001 flare}

\begin{figure}
\begin{center}
\includegraphics*[width=12.0cm]{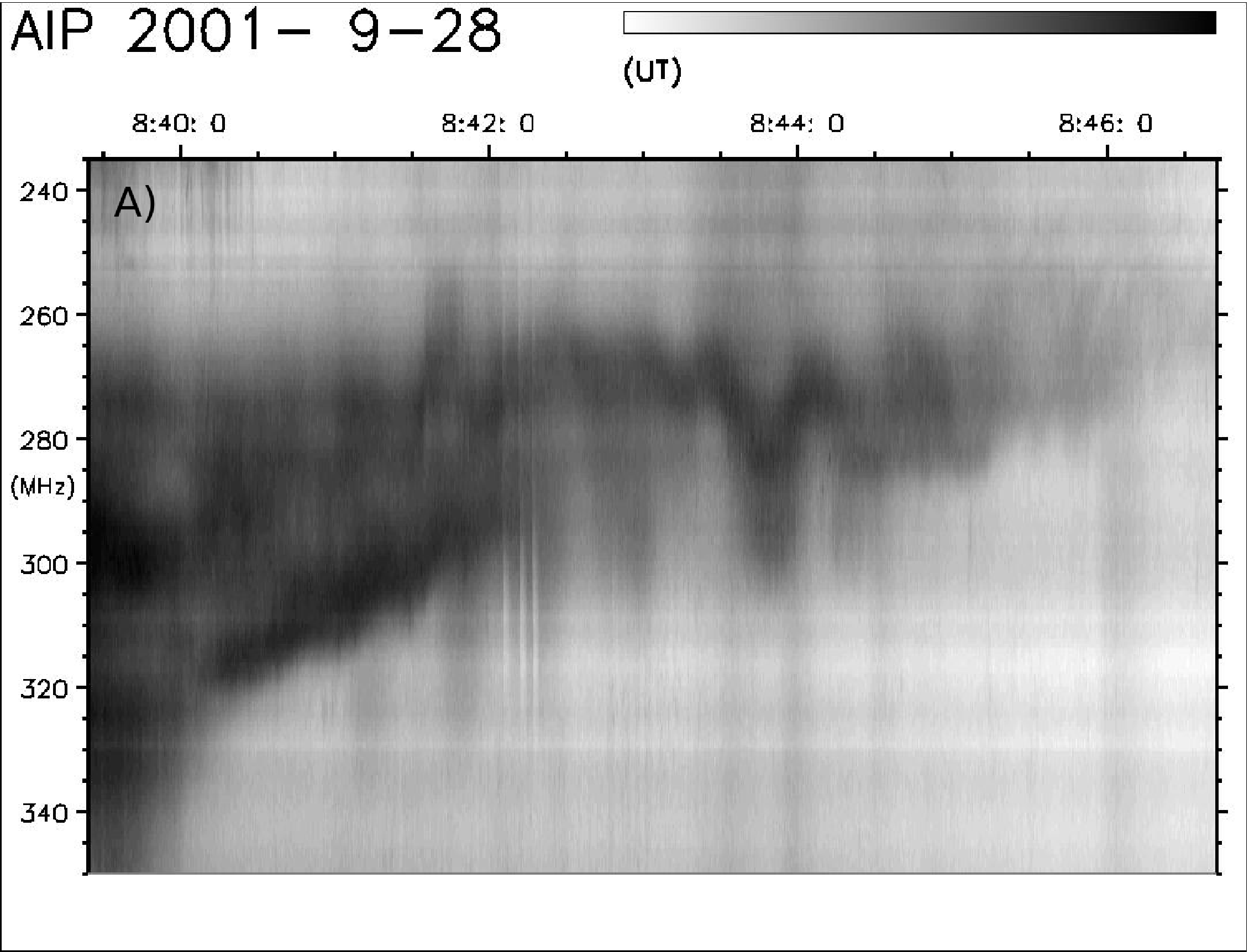}
\includegraphics*[width=18.0cm]{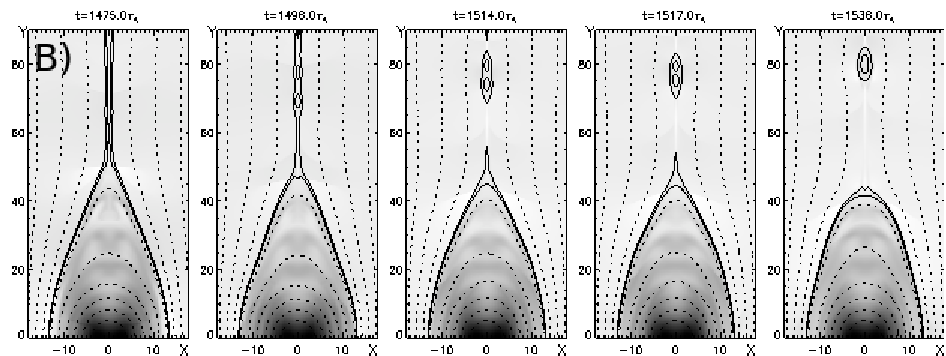}
\end{center}
    \caption{A) Radio spectrum of the 28 September 2001 flare observed by the Potsdam-Tremsdorf radiospectrograph showing
    two DPSs (08:40 - 08:42 UT) that change
    into one frequency-varying DPS (after 08:42 UT). B) Results of numerical simulations showing formation of two plasmoids that merge into one
    oscillating plasmoid, see \cite{2008A&A...477..649B}.}
\label{fig8}
\end{figure}

In the flare current sheet two or even more plasmoids can be formed and merge
into a larger plasmoid (Figure~\ref{fig8}B) \citep{2008A&A...477..649B}. The
merging process generates oscillations of the resulting plasmoid
\citep{1987ApJ...321.1031T}. As presented in Figures 3-5 in the paper by
\citep{2017A&A...602A.122K} oscillations of the merging plasmoid causes
oscillations of the radio emission frequency. Considering the results of all
these simulations we think that the radio bursts in Figure~\ref{fig8}A
corresponds to these processes. On the spectrum, at 08:39 - 08:42 UT we can see
several DPSs that changed after 08:42 UT into one DPS oscillating in the
frequency. This spectrum was observed during the flare starting at 08:10 UT and
ending at 08:50 UT.

\subsection{Type III bursts observed during the 26 September 2011 and 12 February 2010 flares}

\begin{figure}
\begin{center}
\includegraphics*[width=12.0cm]{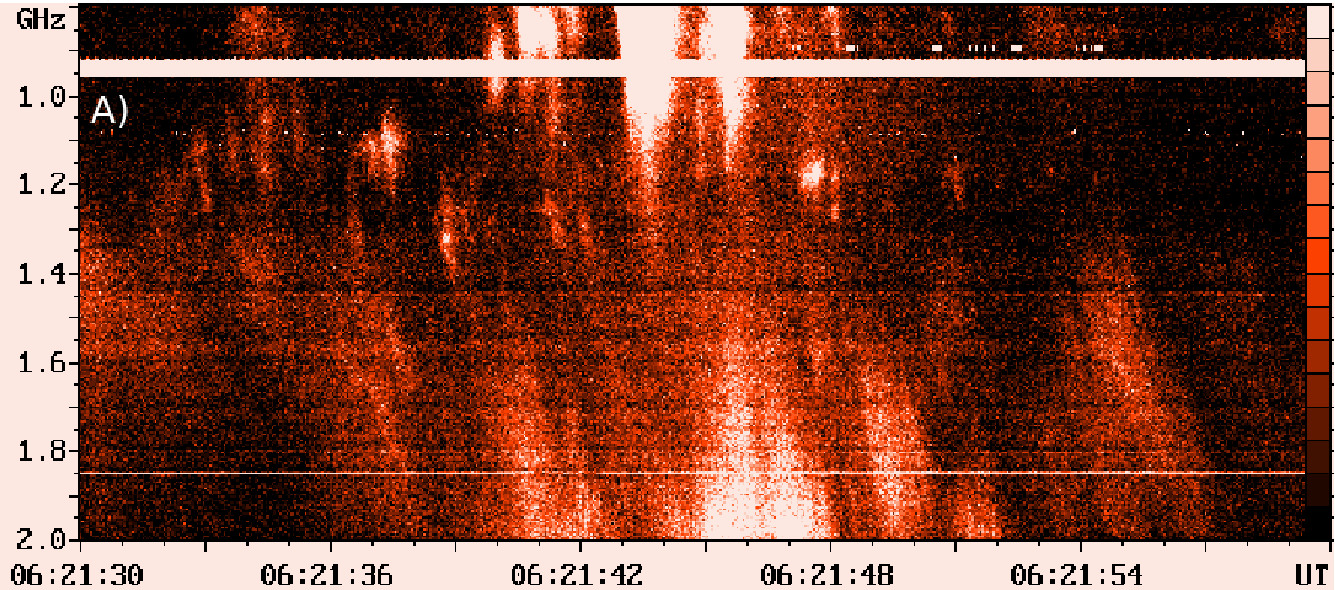}
\includegraphics*[width=10.0cm]{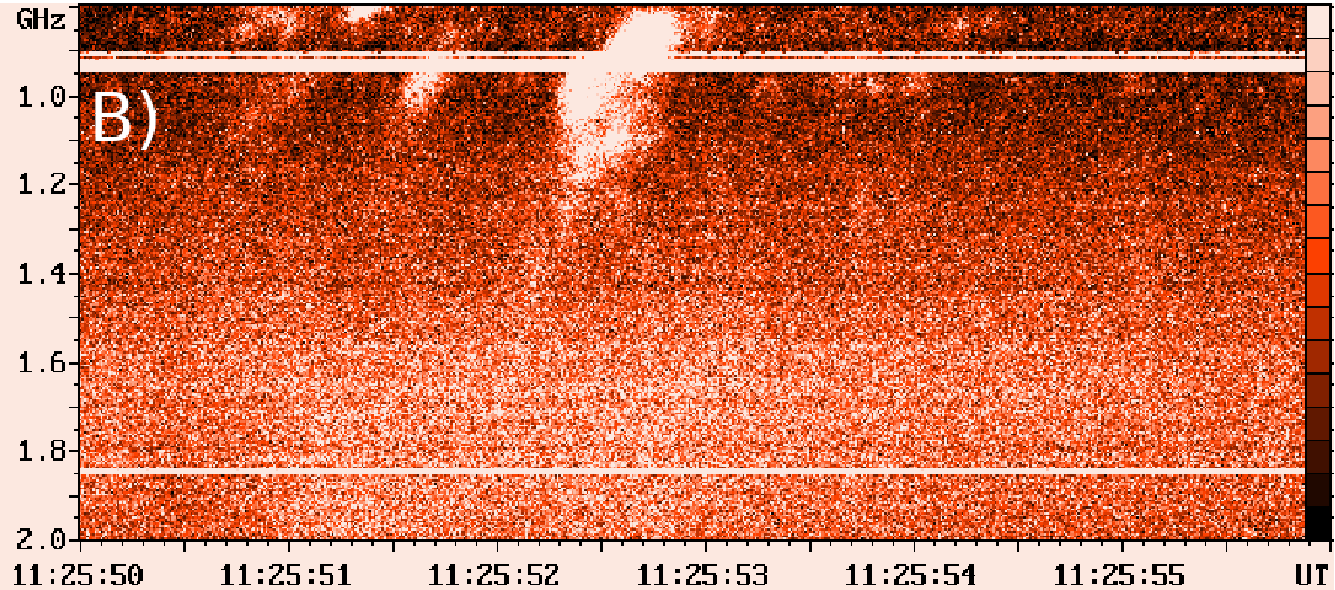}
\includegraphics*[width=6.0cm]{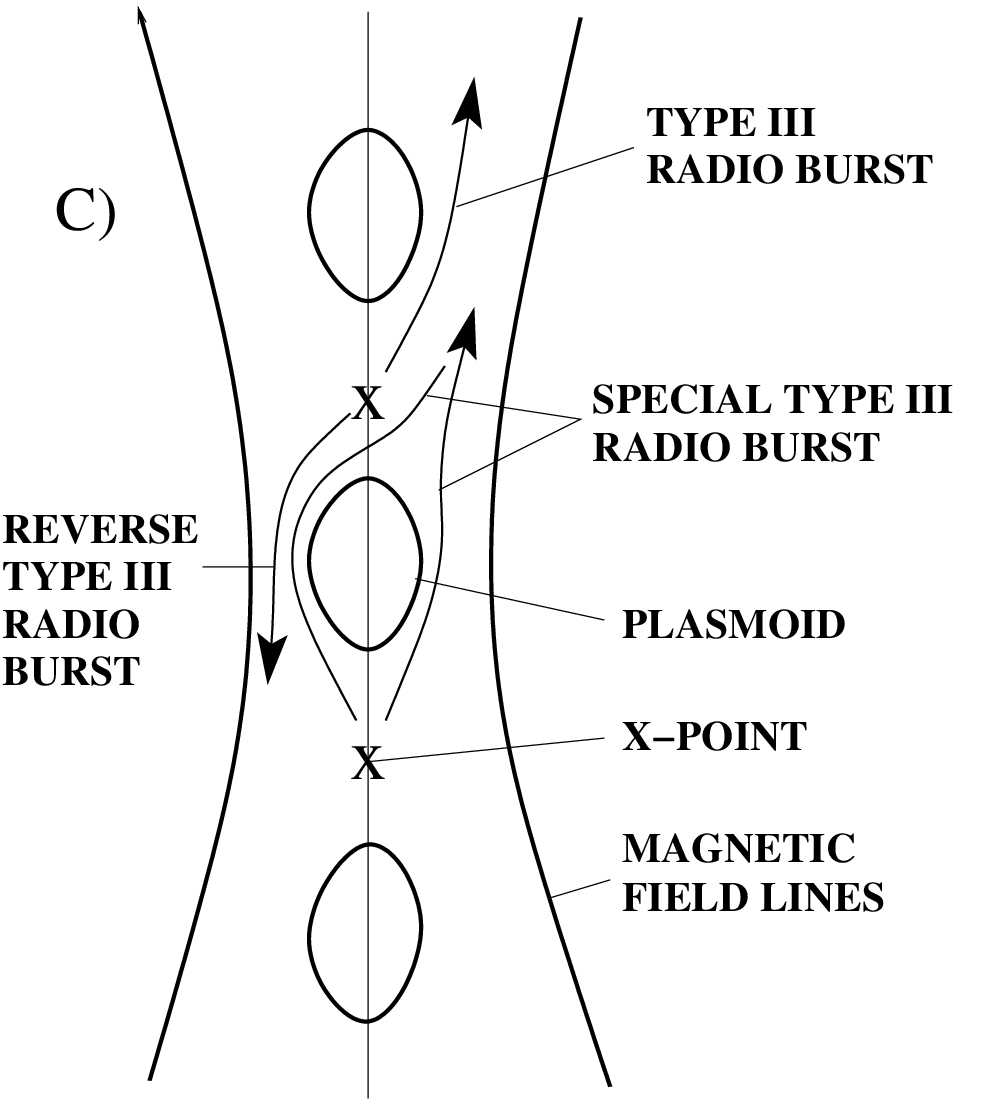}
\end{center}
    \caption{Radio spectra of the 26 September 2011 (A) and 12 February 2010 (B) flares observed by the by the Ond\v{r}ejov radiospectrograph.
    The first one shows pairs of normal and reverse drifting decimetric type III bursts and the second shows an unusual decimetric type III burst.
    C) Scenario for the both bursts explaining them by electron beams.}
\label{fig9}
\end{figure}

Type III bursts are quite common in any solar flares. However, in some flares
special type III bursts are observed. Figure~\ref{fig9} shows two examples of
special type III bursts. In the upper part of this figure pairs of type III
bursts with the normal and reverse (RS) drift are show. This example was
observed during the 26 September 2011 flare (start at 06:21 UT and end at 06:30
UT). The GOES soft X-ray classification is unknown, but during this flare the
hard X-ray emission was observed (\cite{2016SoPh..291.2407T}). These pairs of
type III bursts are explained by the electron beams, accelerated at the X-point
of magnetic reconnection. One beam is propagating upwards (normal type III
burst) and the second one downwards, generating thus the reverse drifting type
III burst. The second example, presented in Figure~\ref{fig9}B, is a unique
type III burst with some hole inside. This special type III burst was observed
during the M9.2 flare (start at 11:25 UT and end 11:32 UT), for more details
about this flare, see \cite{2014Ge&Ae..54..406C}. We propose that in this case
the electron beam accelerated at the X-point of the magnetic reconnection moves
around the plasmoid. To explain the processes generating the both examples of
type III bursts a scenario of assumed processes is shown in Figure~\ref{fig9}C.

\subsection{Decimetric continuum observed during the later phase of the 25 June 2015 flare}

\begin{figure}
\begin{center}
\includegraphics*[width=16.0cm]{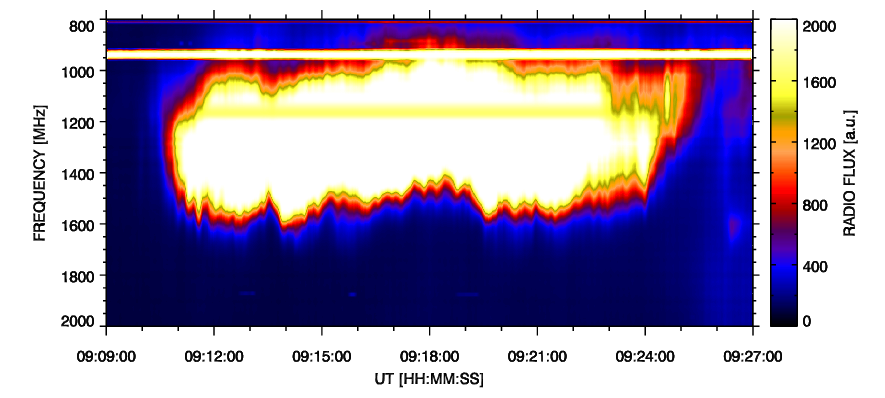}
\includegraphics*[width=6.0cm]{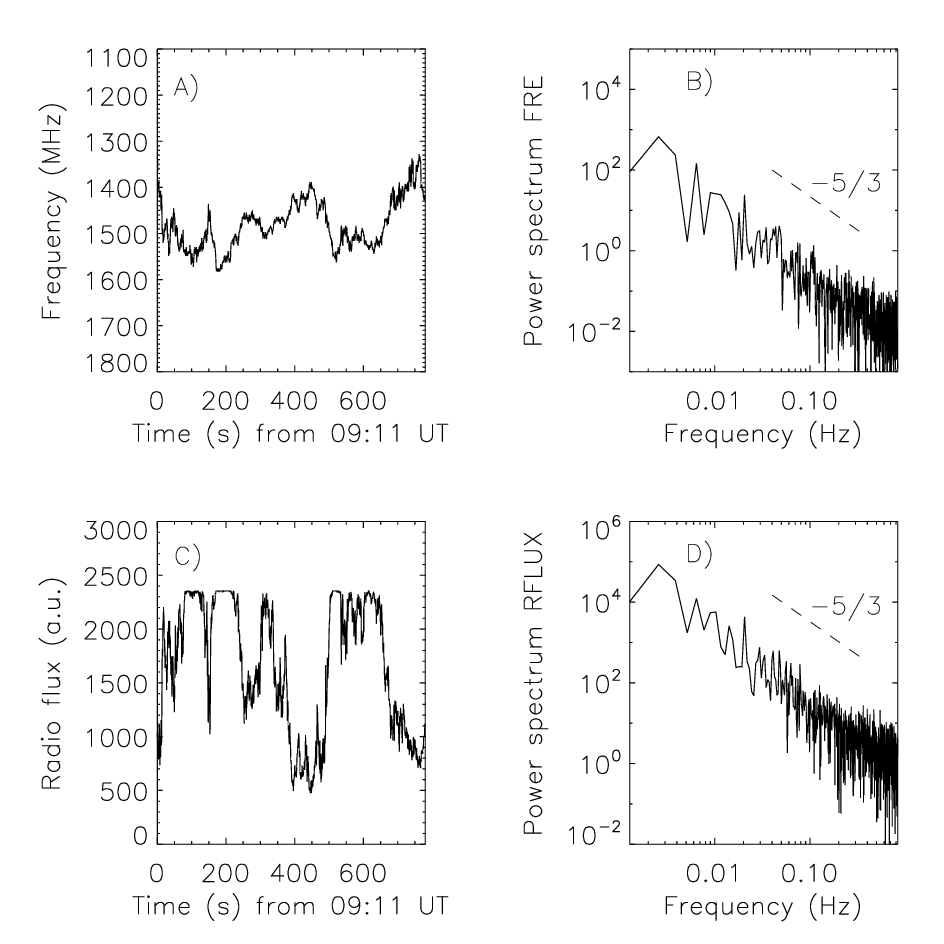}
\includegraphics*[width=8.0cm]{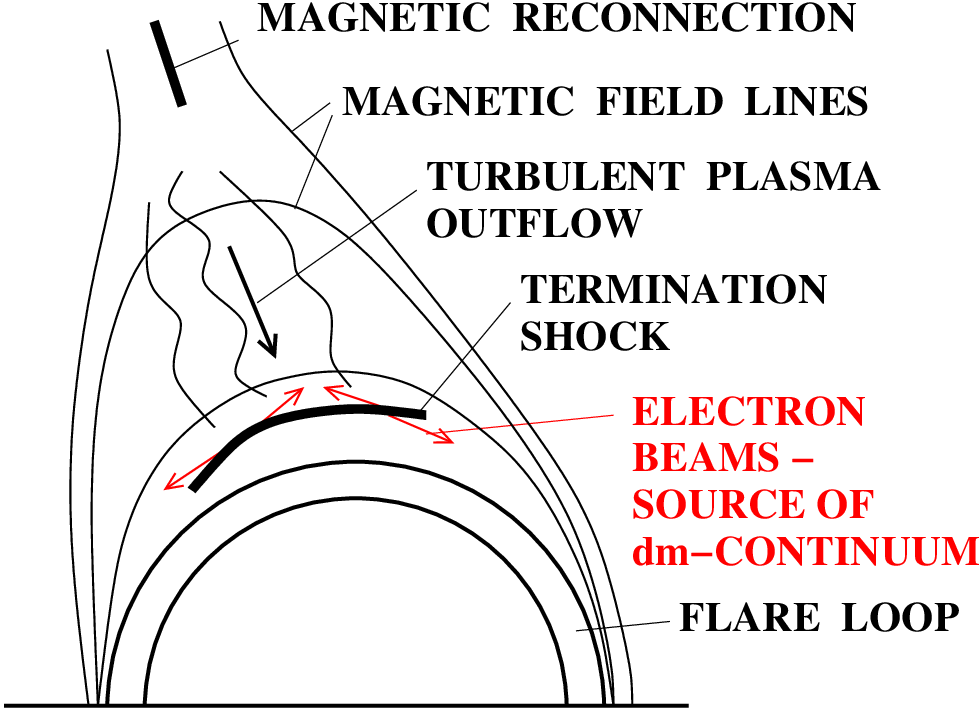}
\end{center}
    \caption{Decimetric continuum observed during the later phase of the 25 June 2015 flare, power spectra (B and D) of frequency variations
    of the high-frequency continuum boundary (A) and radio flux at 1478 MHz (C) in the 09:11-09:24 UT time interval,
    and the decimetric continuum model.}
\label{fig10}
\end{figure}

Figure~\ref{fig10} upper part shows the radio spectrum with the decimetric
continuum observed in the 25 June 2015 flare. In the 09:11-09:24 time interval
we expressed frequency variations of the high-frequency continuum boundary
(Figure~\ref{fig10} bottom part A). Its corresponding power spectrum is in B.
Then we selected the radio flux variations at the frequency in the range of the
continuum boundary variations (C) (at 1478 MHz) and computed its power spectrum
(D). The power-law index in the both cases is close to the Kolmogorov
turbulence index -5/3 that indicates the turbulence in the decimetric continuum
source.

The continuum was observed after the flare impulsive phase when the X-ray flare
emission according to GOES X-ray curves was still enhanced. Therefore, at this
flare phase, in agreement in the standard flare model,  we propose the
following explanation of the presented results, see the schema of the
dm-continuum source in Figure~\ref{fig10}. The plasma outflow from the
reconnection site is in a turbulent state. Owing to this outflow, in front of
the flare loop the termination shock is formed \citep{2009A&A...494..677W}. The
outflow carries the plasma with the magnetic field to the termination shock.
Here, electrons are accelerated by adiabatic reflection in a
quasi-perpendicular shock. This is known as fast-Fermi acceleration or shock
drift acceleration
\citep{1983ApJ...267..837H,1989JGR....9415367K,2011A&A...531A..55V}. Because
the plasma entering the termination shock is in the turbulent state, its
magnetic field changes direction and, therefore, an efficiency of the
acceleration is changed as shown by simulations made by
\cite{2010ApJ...715..406G}. Thus, at the termination shock spatial-spectral
properties of the turbulent plasma outflow are transformed into time-spectral
properties of accelerated electron beams. Then these beams propagate upstream
or downstream of the termination shock and generate radio emission by the
plasma emission mechanism similarly as in the case of herringbones in the type
II bursts. Owing to a high density gradient in the region of continuum
generation the frequency drift of the beam emission is much higher that in the
herringbones case. It is so high that it is not measurable by the used time
resolution of radio spectra. For more details, see \cite{2023K}.

\subsection{Drifting pulsation structure observed during the 10 September 2017 flare together with the EOVSA spatial observation}

\begin{figure}
\begin{center}
\includegraphics*[width=18.0cm]{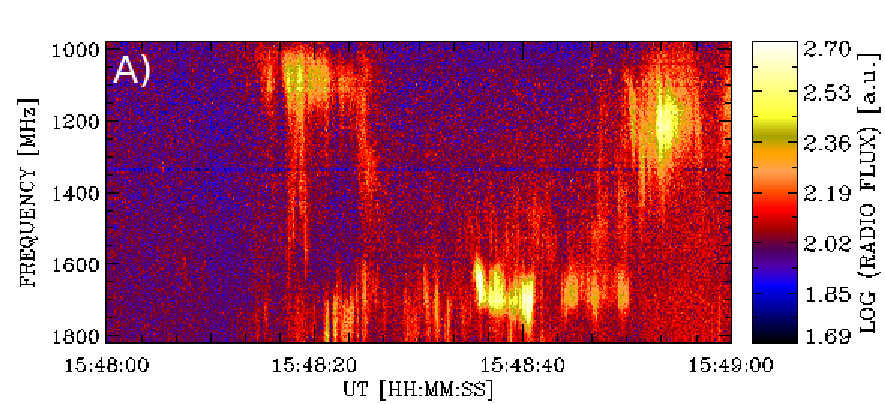}
\includegraphics*[width=18.0cm]{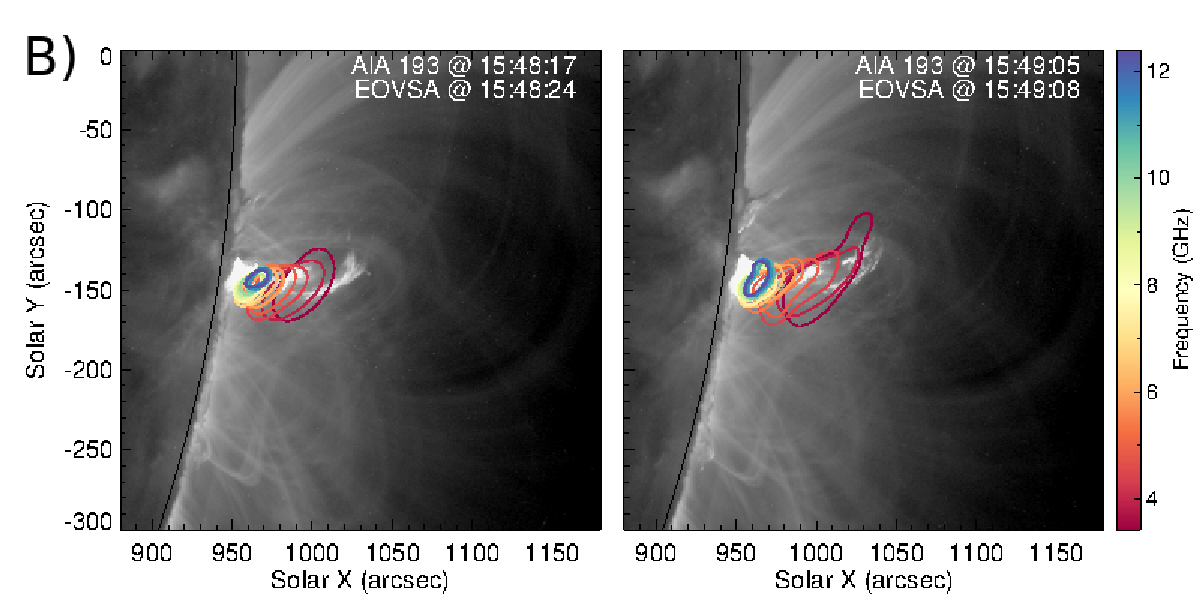}
\end{center}
    \caption{A) Radio spectrum in the 1000-1800 MHz range observed by the Ond\v{r}ejov radiospectrograph
    at the very beginning of the 10 September 2017 flare at 15:48-15:49 UT showing DPS with two branches at 1000-1300 MHz and 1500-1800 MHz.
    B) The EOVSA sources (contours) at 15:48:24 and 15:49:08 UT superimposed on SDO/AIA 193 images
  in the region of a tearing of the ejected filament.}
\label{fig11}
\end{figure}

Positional observations of solar radio bursts in decimetric range, which are
important for the understanding of flare core processes, are not regular.
Nevertheless, new instruments like Expanded Owens Valley Solar Array (EOVSA)
\citep{2013SoPh..288..549G}, Chinese Spectral Radio Heliograph
\citep{2012IAUSS...6E.506L} and Brazilian Decimetre Array
\citep{2007AdSpR..39.1451R} started observations. To show that combined
positional and spectral observations are very important, in Figure~\ref{fig11}A
we show observations of DPS from the X8.2 flare (start at 15:35 UT and end
16:39 UT) simultaneously with EOVSA observations (Figure~\ref{fig11}B).
Although the frequency range of EOVSA observations at this day does not cover
the frequency range of DPS, the observations are very interesting. Namely,
Figure~\ref{fig11} shows that the start and two band structure of DPS coincided
with a tearing of the ejected filament seen by SDO/AIA 193 \AA~observations. At
the same time EOVSA shows a prolongation of sources, especially on the lowest
frequencies ($\sim$ 3.5 GHz). Because later at space of the filament tearing
the flare current sheet was formed \citep{2018ApJ...853L..18Y} we interpret
these observations as follows: When the filament starts to be teared the
current sheet is formed. Inside this sheet the magnetic reconnection starts,
accelerating electrons which are trapped at ends of the broken filament. Thus,
these electrons at these filament ends, having different densities, generate
two bands of DPS. For more details, see \cite{2020ApJ...889...72K}.

\subsection{Schema of bursts based on the presented radio spectra}

\begin{figure}
\begin{center}
\includegraphics*[width=16.0cm]{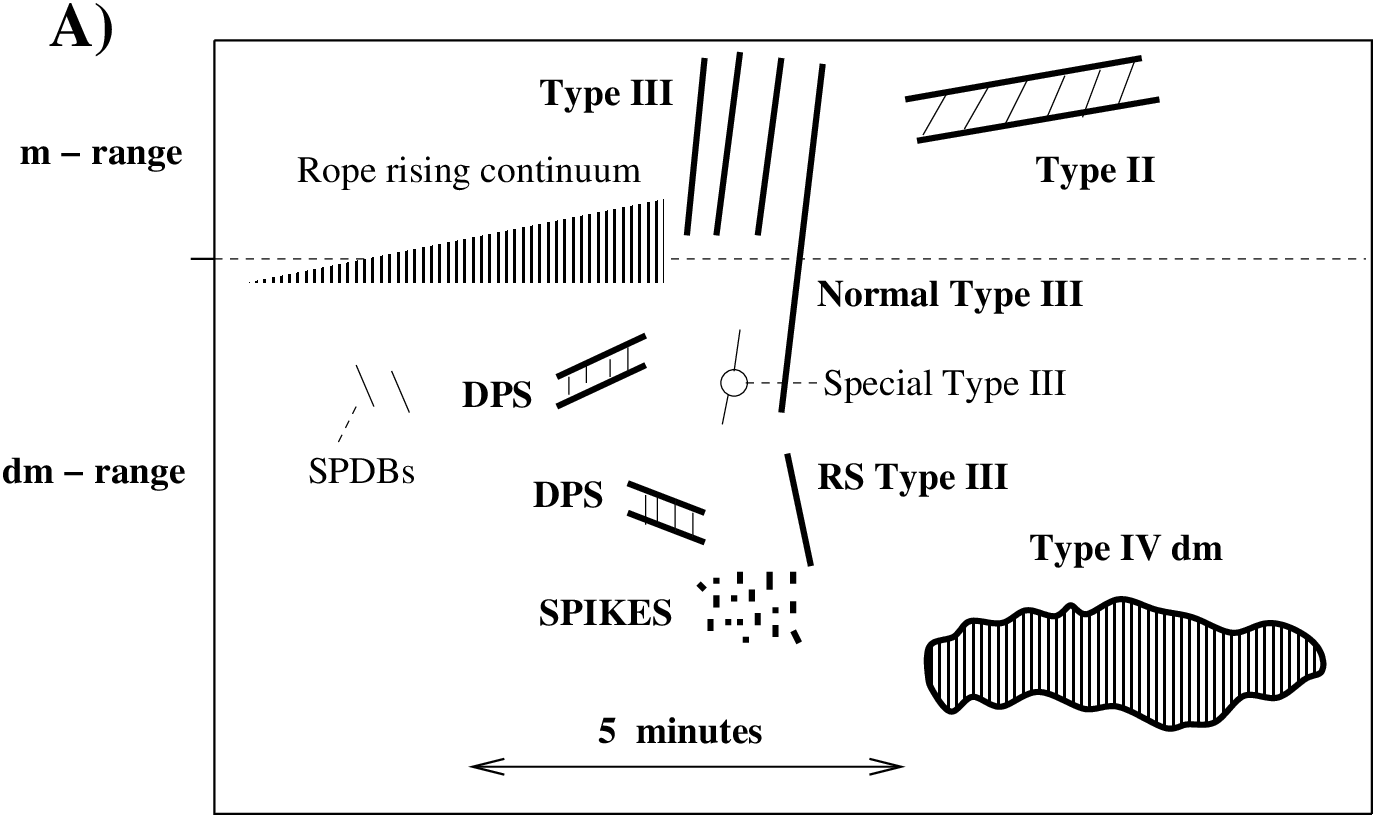}
\includegraphics*[width=16.0cm]{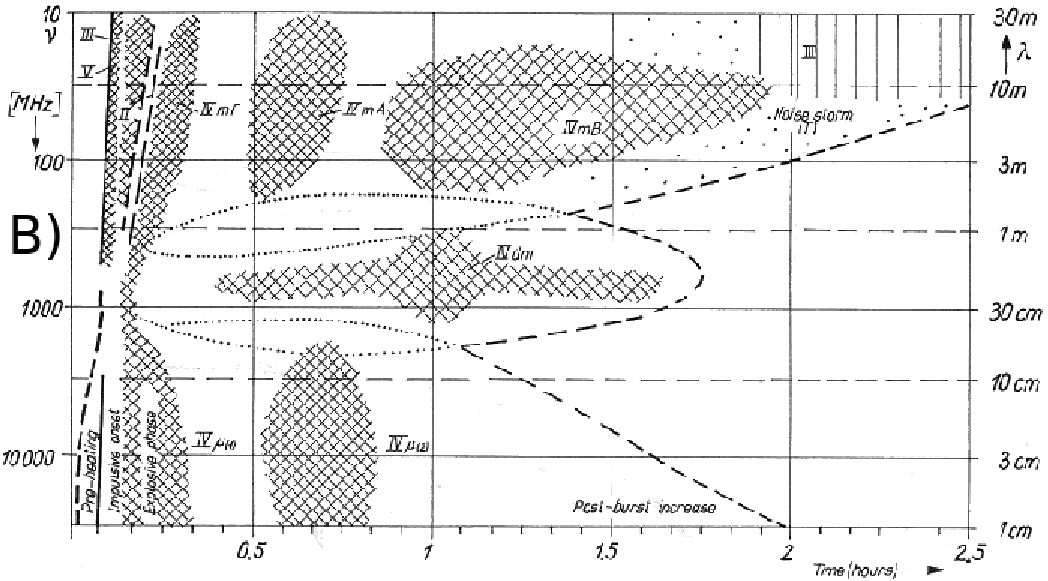}
\end{center}
    \caption{A) Schematic summary of radio bursts observed at the beginning of eruptive flares. The bursts described by thin fonts are very rare.
    B) Standard schema of radio bursts in solar flares according to \cite{1979itsr.book.....K}.}
\label{fig12}
\end{figure}

Considering the frequency range and times of all the presented bursts we
summarize them into the schema shown in Figure~\ref{fig12}A. This schema is
valid for the eruptive flares, especially at their beginnings and in the
decimetric range. While the rope rising continuum, slowly positively drifting
bursts and special type III burst formed around the plasmoid are very rare,
others like DPSs, type III and II bursts, and decimetric continuum are
frequent. For statistical analysis of DPSs and other bursts in the decimetric
bursts, see \cite{2001A&A...375..243J}. The schema starts with the rope rising
continuum or slowly positively drifting bursts, which in many spectra of the
eruptive flares are missing. In most cases the eruptive flares start with DPS
in the decimetric range \citep{2015ApJ...799..126N}. They drift usually towards
lower frequencies. DPS with the positive frequency drift and narrowband
dm-spikes are less frequent. Practically in all eruptive flares we can observe
dm- and m-type III bursts that are in many cases followed by decimetric
continua and type II burst at frequencies in the metric range. For comparison,
in Figure~\ref{fig12}B we added the "standard" schema of radio bursts in solar
flares \citep{1979itsr.book.....K}. Comparing both the schemas we can see that
in the new scheme there are new and rarely observed bursts, that complete the
"standard" schema.

\section{CONCLUSIONS}

We presented examples of some typical and also unusual bursts that were
interpreted using the 3-dimensional model of the eruptive flares and
corresponding numerical simulations. We showed the unusual rope rising
continuum which is associated with the rising magnetic rope at the beginning of
the flare. We also presented slowly positively drifting bursts observed
simultaneously with the bright helical structure inside the rising magnetic
rope. It shows flaring processes in this rope that are not considered in the
standard CSHKP model of the eruptive flares. Namely, in the standard flare
model it is assumed that in the magnetic rope there is a huge electric current,
but its density, which high values are essential for any flaring activity, is
very low due to a large cross-section of the magnetic rope. The primary
energy-release process in the standard flare model is the magnetic reconnection
at the location below the rising magnetic rope.

Typical drifting pulsation structures were presented as signatures of
plasmoids; those with the negative or positive frequency drift as the emission
from the plasmoid moving upwards or downwards in the solar atmosphere. The
radio spectrum corresponding to merging of two plasmoids was also shown. Maybe
that also the unusual zebra, observed at the very early stage of the flare, can
be explained as the emission from the plasmoids distributed in the vertical
flare current sheet.

We also presented the narrowband dm-spikes as a result of the fragmented
magnetic reconnection, a pair of the decimetric type III bursts showing the
electron beams propagating from the acceleration site upwards and downwards in
the solar atmosphere, and an unusual decimetric type III burst probably formed
by the electron beam propagating around the plasmoid. We interpreted decimetric
continua as a radio emission of electron beams accelerated at the termination
shock formed by a turbulent reconnection plasma outflow above flare loops.

We summarized these bursts into a new schema of radio bursts which complete the
previous schema of the radio bursts in solar flares, especially for the
eruptive flares at their beginnings. We showed an importance of observations in
the decimetric range for understanding of the flare processes at the early
stages of eruptive flares. We also presented one example of simultaneous
observations of the drifting pulsation structure together with the positional
observation made by the EOVSA radiointerferometer. However, the observations of
radio source positions in the decimetric range are rare. The example with
positional observations clearly shows how such combined spectral and positional
observations are important. Therefore, new solar-dedicated radio facilities,
working in the decimetric range, like the project of Frequency Agile Solar
Radiotelescope (FASR) that is planned to be an extension of EOVSA
radiointerferometer, are highly desirable.

\acknowledgements We acknowledge support from the project RVO-67985815 and
Grant 22-34841S of the Grant Agency of the Czech Republic.


\begin{thebibliography}{63}
\expandafter\ifx\csname natexlab\endcsname\relax\def\natexlab#1{#1}\fi

\bibitem[{{Akimov} {et~al.}(1996){Akimov}, {Ambro{\v{z}}}, {Belov}, {Berlicki},
  {Chertok}, {Karlick{\'y}}, {Kurt}, {Leikov}, {Litvinenko}, \&
  {Magun}}]{1996SoPh..166..107A}
{Akimov}, V.~V., {Ambro{\v{z}}}, P., {Belov}, A.~V., {et~al.} 1996, \solphys,
  166, 107

\bibitem[{{Aschwanden}(2002)}]{2002SSRv..101....1A}
{Aschwanden}, M.~J. 2002, \ssr, 101, 1

\bibitem[{{Aulanier} {et~al.}(2012){Aulanier}, {Janvier}, \&
  {Schmieder}}]{2012A&A...543A.110A}
{Aulanier}, G., {Janvier}, M., \& {Schmieder}, B. 2012, \aap, 543, A110

\bibitem[{{Aurass} \& {Mann}(2004)}]{2004ApJ...615..526A}
{Aurass}, H. \& {Mann}, G. 2004, \apj, 615, 526

\bibitem[{{Aurass} {et~al.}(2011){Aurass}, {Mann}, {Zlobec}, \&
  {Karlick{\'y}}}]{2011ApJ...730...57A}
{Aurass}, H., {Mann}, G., {Zlobec}, P., \& {Karlick{\'y}}, M. 2011, \apj, 730,
  57

\bibitem[{{B{\'a}rta} {et~al.}(2008{\natexlab{a}}){B{\'a}rta}, {Karlick{\'y}},
  \& {{\v Z}emli{\v c}ka}}]{2008SoPh..253..173B}
{B{\'a}rta}, M., {Karlick{\'y}}, M., \& {{\v Z}emli{\v c}ka}, R.
  2008{\natexlab{a}}, \solphys, 253, 173

\bibitem[{{B{\'a}rta} {et~al.}(2008{\natexlab{b}}){B{\'a}rta}, {Vr{\v s}nak},
  \& {Karlick{\'y}}}]{2008A&A...477..649B}
{B{\'a}rta}, M., {Vr{\v s}nak}, B., \& {Karlick{\'y}}, M. 2008{\natexlab{b}},
  \aap, 477, 649

\bibitem[{{Brown}(1971)}]{1971SoPh...18..489B}
{Brown}, J.~C. 1971, \solphys, 18, 489

\bibitem[{{Carmichael}(1964)}]{1964NASSP..50..451C}
{Carmichael}, H. 1964, NASA Special Publication, 50, 451

\bibitem[{{Chernov}(2011)}]{2011fssr.book.....C}
{Chernov}, G. 2011, {Fine Structure of Solar Radio Bursts, Berlin: Springer}

\bibitem[{{Chernov} {et~al.}(2014){Chernov}, {Fomichev}, {Gorgutsa}, {Markeev},
  {Sobolev}, {Hillaris}, \& {Alissandrakis}}]{2014Ge&Ae..54..406C}
{Chernov}, G.~P., {Fomichev}, V.~V., {Gorgutsa}, R.~V., {et~al.} 2014,
  Geomagnetism and Aeronomy, 54, 406

\bibitem[{{Dahlin} {et~al.}(2014){Dahlin}, {Drake}, \&
  {Swisdak}}]{2014PhPl...21i2304D}
{Dahlin}, J.~T., {Drake}, J.~F., \& {Swisdak}, M. 2014, Physics of Plasmas, 21,
  092304

\bibitem[{{F{\'a}rn{\'{\i}}k} {et~al.}(2003){F{\'a}rn{\'{\i}}k}, {Hudson},
  {Karlick{\'y}}, \& {Kosugi}}]{2003A&A...399.1159F}
{F{\'a}rn{\'{\i}}k}, F., {Hudson}, H.~S., {Karlick{\'y}}, M., \& {Kosugi}, T.
  2003, \aap, 399, 1159

\bibitem[{{Fletcher} {et~al.}(2011){Fletcher}, {Dennis}, {Hudson}, {Krucker},
  {Phillips}, {Veronig}, {Battaglia}, {Bone}, {Caspi}, {Chen}, {Gallagher},
  {Grigis}, {Ji}, {Liu}, {Milligan}, \& {Temmer}}]{2011SSRv..159...19F}
{Fletcher}, L., {Dennis}, B.~R., {Hudson}, H.~S., {et~al.} 2011, \ssr, 159, 19

\bibitem[{{Gary} {et~al.}(2013){Gary}, {Fleishman}, \&
  {Nita}}]{2013SoPh..288..549G}
{Gary}, D.~E., {Fleishman}, G.~D., \& {Nita}, G.~M. 2013, \solphys, 288, 549

\bibitem[{{Guo} \& {Giacalone}(2010)}]{2010ApJ...715..406G}
{Guo}, F. \& {Giacalone}, J. 2010, \apj, 715, 406

\bibitem[{{Hirayama}(1974)}]{1974SoPh...34..323H}
{Hirayama}, T. 1974, \solphys, 34, 323

\bibitem[{{Holman} \& {Pesses}(1983)}]{1983ApJ...267..837H}
{Holman}, G.~D. \& {Pesses}, M.~E. 1983, \apj, 267, 837

\bibitem[{{Huang} {et~al.}(2017){Huang}, {Comisso}, \&
  {Bhattacharjee}}]{2017ApJ...849...75H}
{Huang}, Y.-M., {Comisso}, L., \& {Bhattacharjee}, A. 2017, \apj, 849, 75

\bibitem[{{Janvier} {et~al.}(2014){Janvier}, {Aulanier}, {Bommier},
  {Schmieder}, {D{\'e}moulin}, \& {Pariat}}]{2014ApJ...788...60J}
{Janvier}, M., {Aulanier}, G., {Bommier}, V., {et~al.} 2014, \apj, 788, 60

\bibitem[{{Ji{\v r}i{\v c}ka} \& {Karlick{\'y}}(2008)}]{2008SoPh..253...95J}
{Ji{\v r}i{\v c}ka}, K. \& {Karlick{\'y}}, M. 2008, \solphys, 253, 95

\bibitem[{{Ji{\v r}i{\v c}ka} {et~al.}(2001){Ji{\v r}i{\v c}ka},
  {Karlick{\'y}}, {M{\'e}sz{\'a}rosov{\'a}}, \& {Sn{\'{\i}}{\v
  z}ek}}]{2001A&A...375..243J}
{Ji{\v r}i{\v c}ka}, K., {Karlick{\'y}}, M., {M{\'e}sz{\'a}rosov{\'a}}, H., \&
  {Sn{\'{\i}}{\v z}ek}, V. 2001, \aap, 375, 243

\bibitem[{{Ji\v{r}i\v{c}ka} {et~al.}(1993){Ji\v{r}i\v{c}ka}, {Karlick\'y},
  {Kepka}, \& {Tlamicha}}]{1993SoPh..147..203J}
{Ji\v{r}i\v{c}ka}, K., {Karlick\'y}, M., {Kepka}, O., \& {Tlamicha}, A. 1993,
  \solphys, 147, 203

\bibitem[{{Karlick{\'y}}(2008)}]{2008ApJ...674.1211K}
{Karlick{\'y}}, M. 2008, \apj, 674, 1211

\bibitem[{{Karlick{\'y}}(2017)}]{2017A&A...602A.122K}
{Karlick{\'y}}, M. 2017, \aap, 602, A122

\bibitem[{{Karlick{\'y}}(2023)}]{2023K}
{Karlick{\'y}}, M. 2023, \solphys, accepted

\bibitem[{{Karlick{\'y}} {et~al.}(2020{\natexlab{a}}){Karlick{\'y}}, {Chen},
  {Gary}, {Ka{\v{s}}parov{\'a}}, \& {Ryb{\'a}k}}]{2020ApJ...889...72K}
{Karlick{\'y}}, M., {Chen}, B., {Gary}, D.~E., {Ka{\v{s}}parov{\'a}}, J., \&
  {Ryb{\'a}k}, J. 2020{\natexlab{a}}, \apj, 889, 72

\bibitem[{{Karlick{\'y}} {et~al.}(2002){Karlick{\'y}}, {F{\'a}rn{\'{\i}}k}, \&
  {M{\'e}sz{\'a}rosov{\'a}}}]{2002A&A...395..677K}
{Karlick{\'y}}, M., {F{\'a}rn{\'{\i}}k}, F., \& {M{\'e}sz{\'a}rosov{\'a}}, H.
  2002, \aap, 395, 677

\bibitem[{{Karlick{\'y}} {et~al.}(2000){Karlick{\'y}}, {Ji{\v{r}}i{\v{c}}ka},
  \& {Sobotka}}]{2000SoPh..195..165K}
{Karlick{\'y}}, M., {Ji{\v{r}}i{\v{c}}ka}, K., \& {Sobotka}, M. 2000, \solphys,
  195, 165

\bibitem[{{Karlick\'y} \& {Jungwirth}(1989)}]{1989SoPh..124..319K}
{Karlick\'y}, M. \& {Jungwirth}, K. 1989, \solphys, 124, 319

\bibitem[{{Karlick{\'y}} {et~al.}(2020{\natexlab{b}}){Karlick{\'y}},
  {Ka{\v{s}}parov{\'a}}, \& {Sych}}]{2020ApJ...888...18K}
{Karlick{\'y}}, M., {Ka{\v{s}}parov{\'a}}, J., \& {Sych}, R.
  2020{\natexlab{b}}, \apj, 888, 18

\bibitem[{{Karlick{\'y}} {et~al.}(1996){Karlick{\'y}}, {Sobotka}, \&
  {Ji{\v{r}}i{\v{c}}ka}}]{1996SoPh..168..375K}
{Karlick{\'y}}, M., {Sobotka}, M., \& {Ji{\v{r}}i{\v{c}}ka}, K. 1996, \solphys,
  168, 375

\bibitem[{{Kliem} {et~al.}(2000){Kliem}, {Karlick{\'y}}, \&
  {Benz}}]{2000A&A...360..715K}
{Kliem}, B., {Karlick{\'y}}, M., \& {Benz}, A.~O. 2000, \aap, 360, 715

\bibitem[{{Kliem} {et~al.}(2010){Kliem}, {Linton}, {T{\"o}r{\"o}k}, \&
  {Karlick{\'y}}}]{2010SoPh..266...91K}
{Kliem}, B., {Linton}, M.~G., {T{\"o}r{\"o}k}, T., \& {Karlick{\'y}}, M. 2010,
  \solphys, 266, 91

\bibitem[{{Kliem} \& {T{\"o}r{\"o}k}(2006)}]{2006PhRvL..96y5002K}
{Kliem}, B. \& {T{\"o}r{\"o}k}, T. 2006, \prl, 96, 255002

\bibitem[{{Kopp} \& {Pneuman}(1976)}]{1976SoPh...50...85K}
{Kopp}, R.~A. \& {Pneuman}, G.~W. 1976, \solphys, 50, 85

\bibitem[{{Kosovichev} \& {Zharkova}(1998)}]{1998Natur.393..317K}
{Kosovichev}, A.~G. \& {Zharkova}, V.~V. 1998, \nat, 393, 317

\bibitem[{{Krauss-Varban} \& {Wu}(1989)}]{1989JGR....9415367K}
{Krauss-Varban}, D. \& {Wu}, C.~S. 1989, \jgr, 94, 15367

\bibitem[{{Krucker} {et~al.}(2008){Krucker}, {Battaglia}, {Cargill},
  {Fletcher}, {Hudson}, {MacKinnon}, {Masuda}, {Sui}, {Tomczak}, {Veronig},
  {Vlahos}, \& {White}}]{2008A&ARv..16..155K}
{Krucker}, S., {Battaglia}, M., {Cargill}, P.~J., {et~al.} 2008, \aapr, 16, 155

\bibitem[{{Krueger}(1979)}]{1979itsr.book.....K}
{Krueger}, A. 1979, {Introduction to solar radio astronomy and radio physics,
  Geophysics and Astrophysics Monographs, Dordrecht: Reidel, 1979, p.115}

\bibitem[{{Li} {et~al.}(2012){Li}, {Yan}, {Wang}, \&
  {Liu}}]{2012IAUSS...6E.506L}
{Li}, S., {Yan}, Y., {Wang}, W., \& {Liu}, D. 2012, IAU Special Session, 6,
  E5.06

\bibitem[{{M{\'e}sz{\'a}rosov{\'a}} {et~al.}(2006){M{\'e}sz{\'a}rosov{\'a}},
  {Karlick{\'y}}, {Ryb{\'a}k}, {F{\'a}rn{\'{\i}}k}, \& {Ji{\v r}i{\v
  c}ka}}]{2006A&A...460..865M}
{M{\'e}sz{\'a}rosov{\'a}}, H., {Karlick{\'y}}, M., {Ryb{\'a}k}, J.,
  {F{\'a}rn{\'{\i}}k}, F., \& {Ji{\v r}i{\v c}ka}, K. 2006, \aap, 460, 865

\bibitem[{{M{\'e}sz{\'a}rosov{\'a}} {et~al.}(2003){M{\'e}sz{\'a}rosov{\'a}},
  {Veronig}, {Zlobec}, \& {Karlick{\'y}}}]{2003A&A...407.1115M}
{M{\'e}sz{\'a}rosov{\'a}}, H., {Veronig}, A., {Zlobec}, P., \& {Karlick{\'y}},
  M. 2003, \aap, 407, 1115

\bibitem[{{Mitra} {et~al.}(2018){Mitra}, {Joshi}, {Prasad}, {Veronig}, \&
  {Bhattacharyya}}]{2018ApJ...869...69M}
{Mitra}, P.~K., {Joshi}, B., {Prasad}, A., {Veronig}, A.~M., \&
  {Bhattacharyya}, R. 2018, \apj, 869, 69

\bibitem[{{Moreton}(1960)}]{1960AJ.....65U.494M}
{Moreton}, G.~E. 1960, \aj, 65, 494

\bibitem[{{Nakariakov} {et~al.}(2016){Nakariakov}, {Pilipenko}, {Heilig},
  {Jel{\'{\i}}nek}, {Karlick{\'y}}, {Klimushkin}, {Kolotkov}, {Lee},
  {Nistic{\`o}}, {Van Doorsselaere}, {Verth}, \&
  {Zimovets}}]{2016SSRv..200...75N}
{Nakariakov}, V.~M., {Pilipenko}, V., {Heilig}, B., {et~al.} 2016, \ssr, 200,
  75

\bibitem[{{Nelson} \& {Melrose}(1985)}]{1985srph.book..333N}
{Nelson}, G.~J. \& {Melrose}, D.~B. 1985, {Type II bursts.}, ed. D.~J. {McLean}
  \& N.~R. {Labrum}, 333--359

\bibitem[{{Nishizuka} {et~al.}(2015){Nishizuka}, {Karlick{\'y}}, {Janvier}, \&
  {B{\'a}rta}}]{2015ApJ...799..126N}
{Nishizuka}, N., {Karlick{\'y}}, M., {Janvier}, M., \& {B{\'a}rta}, M. 2015,
  \apj, 799, 126

\bibitem[{{Odstr\v{c}il} \& {Karlick\'y}(1997)}]{1997A&A...326.1252O}
{Odstr\v{c}il}, D. \& {Karlick\'y}, M. 1997, \aap, 326, 1252

\bibitem[{{Priest} \& {Forbes}(2002)}]{2002A&ARv..10..313P}
{Priest}, E.~R. \& {Forbes}, T.~G. 2002, \aapr, 10, 313

\bibitem[{{Ramesh} {et~al.}(2007){Ramesh}, {Sawant}, {Cecatto}, {Faria},
  {Fernandes}, {Kathiravan}, \& {Suryanarayana}}]{2007AdSpR..39.1451R}
{Ramesh}, R., {Sawant}, H.~S., {Cecatto}, J.~R., {et~al.} 2007, Advances in
  Space Research, 39, 1451

\bibitem[{{Sato} {et~al.}(2018){Sato}, {Jakowski}, {Berdermann},
  {Ji\v{r}i\v{c}ka}, \& {He{\ss}elbarth}}]{2018AGUFMSA34A..04S}
{Sato}, H., {Jakowski}, N., {Berdermann}, J., {Ji\v{r}i\v{c}ka}, K., \&
  {He{\ss}elbarth}, A. 2018, in AGU Fall Meeting Abstracts, Vol. 2018,
  SA34A--04

\bibitem[{{Schrijver}(2009)}]{2009AdSpR..43..739S}
{Schrijver}, C.~J. 2009, Advances in Space Research, 43, 739

\bibitem[{{Sturrock}(1966)}]{1966Natur.211..695S}
{Sturrock}, P.~A. 1966, \nat, 211, 695

\bibitem[{{Suzuki} \& {Dulk}(1985)}]{1985srph.book..289S}
{Suzuki}, S. \& {Dulk}, G.~A. 1985, {Bursts of type III and type V.}, ed. D.~J.
  {McLean} \& N.~R. {Labrum}, 289--332

\bibitem[{{\v{S}vestka} {et~al.}(1992){\v{S}vestka}, {Jackson}, \&
  {Machado}}]{1992LNP...399.....S}
{\v{S}vestka}, Z., {Jackson}, B.~V., \& {Machado}, M.~E., eds. 1992, Lecture
  Notes in Physics, Berlin Springer Verlag, Vol. 399, {Eruptive Solar Flares}

\bibitem[{{Tajima} {et~al.}(1987){Tajima}, {Sakai}, {Nakajima}, {Kosugi},
  {Brunel}, \& {Kundu}}]{1987ApJ...321.1031T}
{Tajima}, T., {Sakai}, J., {Nakajima}, H., {et~al.} 1987, \apj, 321, 1031

\bibitem[{{Tan} {et~al.}(2016){Tan}, {Karlick{\'y}}, {M{\'e}sz{\'a}rosov{\'a}},
  {Kashapova}, {Huang}, {Yan}, \& {Kontar}}]{2016SoPh..291.2407T}
{Tan}, B.~L., {Karlick{\'y}}, M., {M{\'e}sz{\'a}rosov{\'a}}, H., {et~al.} 2016,
  \solphys, 291, 2407

\bibitem[{{Thompson} {et~al.}(1998){Thompson}, {Plunkett}, {Gurman}, {Newmark},
  {St. Cyr}, \& {Michels}}]{1998GeoRL..25.2465T}
{Thompson}, B.~J., {Plunkett}, S.~P., {Gurman}, J.~B., {et~al.} 1998, \grl, 25,
  2465

\bibitem[{{Vandas} \& {Karlick{\'y}}(2011)}]{2011A&A...531A..55V}
{Vandas}, M. \& {Karlick{\'y}}, M. 2011, \aap, 531, A55

\bibitem[{{Warmuth} {et~al.}(2009){Warmuth}, {Mann}, \&
  {Aurass}}]{2009A&A...494..677W}
{Warmuth}, A., {Mann}, G., \& {Aurass}, H. 2009, \aap, 494, 677

\bibitem[{{Yan} {et~al.}(2018){Yan}, {Yang}, {Xue}, {Mei}, {Kong}, {Wang}, \&
  {Li}}]{2018ApJ...853L..18Y}
{Yan}, X.~L., {Yang}, L.~H., {Xue}, Z.~K., {et~al.} 2018, \apjl, 853, L18

\bibitem[{{Zemanov{\'a}} {et~al.}(2020){Zemanov{\'a}}, {Karlick{\'y}},
  {Ka{\v{s}}parov{\'a}}, \& {Dud{\'\i}k}}]{2020ApJ...905..111Z}
{Zemanov{\'a}}, A., {Karlick{\'y}}, M., {Ka{\v{s}}parov{\'a}}, J., \&
  {Dud{\'\i}k}, J. 2020, \apj, 905, 111

\end{thebibliography}

\newpage

\end{document}